%% file: ISSTA_Main.tex
\AtBeginDocument{%
  \providecommand\BibTeX{{%
    \normalfont B\kern-0.5em{\scshape i\kern-0.25em b}\kern-0.8em\TeX}}}

\documentclass[sigconf,screen]{acmart}

\setcopyright{acmlicensed}
\acmPrice{15.00}
\acmDOI{10.1145/3597926.3598117}
\acmYear{2023}
\copyrightyear{2023}
\acmSubmissionID{issta23main-p285-p}
\acmISBN{979-8-4007-0221-1/23/07}
\acmConference[ISSTA '23]{Proceedings of the 32nd ACM SIGSOFT International Symposium on Software Testing and Analysis}{July 17--21, 2023}{Seattle, WA, USA}
\acmBooktitle{Proceedings of the 32nd ACM SIGSOFT International Symposium on Software Testing and Analysis (ISSTA '23), July 17--21, 2023, Seattle, WA, USA}
\received{2023-02-16}
\received[accepted]{2023-05-03}

\usepackage{amsmath,amsfonts}
\usepackage{multirow}
\usepackage{algorithmic} 

\usepackage{graphicx}
\usepackage{textcomp}
\usepackage{xcolor}
\usepackage{booktabs}
\usepackage{adjustbox}
\usepackage{subcaption}
\usepackage{float}
\usepackage{graphicx}
\usepackage{caption}
\usepackage{subcaption}
\usepackage{tcolorbox}

\usepackage{hyperref}

\usepackage[linesnumbered,algoruled,boxed,lined]{algorithm2e}
\usepackage{xcolor}
\usepackage{subcaption}
\usepackage[utf8]{inputenc}
\usepackage{flushend}

\SetKwRepeat{Do}{do}{while}
\let\oldnl\nl
\newcommand{\nonl}{\renewcommand{\nl}{\let\nl\oldnl}}
\begin{document}

\title{Applying and Extending the Delta Debugging Algorithm for Elevator Dispatching Algorithms (Experience Paper)}

\author{Pablo Valle}
\email{pablo.valle@alumni.mondragon.edu}
\orcid{0000-0002-0588-316X}
\affiliation{%
  \institution{Mondragon University}
  \streetaddress{Goiru 2}
  \city{Mondragon}
  \state{Gipuzkoa}
  \country{Spain}
  \postcode{20500}
}

\author{Aitor Arrieta}
\email{aarrieta@mondragon.edu}
\orcid{0000-0001-7507-5080}
\affiliation{%
  \institution{Mondragon University}
  \streetaddress{Goiru 2}
  \city{Mondragon}
  \state{Gipuzkoa}
  \country{Spain}
  \postcode{20500}
}

\author{Maite Arratibel}
\email{marratibel@orona-group.com}
\orcid{0000-0002-9880-0764}
\affiliation{%
  \institution{Orona}
  \city{Hernani}
  \state{Gipuzkoa}
  \country{Spain}
}

\renewcommand{\shortauthors}{Pablo Valle, Aitor Arrieta and Maite Arratibel}

\begin{abstract}
Elevator systems are one kind of Cyber-Physical Systems (CPSs), and as such, test cases are usually complex and long in time. This is mainly because realistic test scenarios are employed (e.g., for testing elevator dispatching algorithms, typically a full day of passengers traveling through a system of elevators is used). However, in such a context, when needing to reproduce a failure, it is of high benefit to provide the minimal test input to the software developers. This way, analyzing and trying to localize the root-cause of the failure is easier and more agile. Delta debugging has been found to be an efficient technique to reduce failure-inducing test inputs. In this paper, we enhance this technique by first monitoring the environment at which the CPS operates as well as its physical states. With the monitored information, we search for stable states of the CPS during the execution of the simulation. In a second step, we use such identified stable states to help the delta debugging algorithm isolate the failure-inducing test inputs more efficiently. 

We report our experience of applying our approach into an industrial elevator dispatching algorithm. An empirical evaluation carried out with real operational data from a real installation of elevators suggests that the proposed environment-wise delta debugging algorithm is between 1.3 to 1.8 times faster than the traditional delta debugging, while producing a larger reduction in the failure-inducing test inputs. The results provided by the different implemented delta debugging algorithm versions are qualitatively assessed with domain experts. This assessment provides new insights and lessons learned, such as, potential applications of the delta debugging algorithm beyond debugging.
\end{abstract}






\maketitle

\section{Introduction}

\input{introduction}

\section{Industrial Case Study} \label{sec:industrialCaseStudy}
\input{casestudy}

\section{Approach}\label{sec:Approach}
\input{approach}

\section{Empirical Evaluation}\label{sec:Evaluation}
\input{evaluation}

\section{Lessons Learned}\label{sec:LessonsLearned}
\input{lessons}

\section{Related Work}\label{sec:RelatedWork}
\input{relatedWork}

\section{Conclusion and Future Work}\label{sec:Conclusions}
\input{conclusion}

\section*{Acknowledgments}
 Project supported by a 2021 Leonardo Grant for Researchers and Cultural Creators, BBVA Foundation. The BBVA Foundation is not responsible for the opinions, comments and contents included in the project and/or the results derived from it, which are the total and absolute responsibility of their authors. Aitor Arrieta is part of the Software and Systems Engineering research group of Mondragon Unibertsitatea (IT1519-22), supported by the Department of Education, Universities and Research of the Basque Country.

\bibliographystyle{ACM-Reference-Format}
\bibliography{bibliografia}
\end{document}

%% file: introduction.tex
A system of elevators is a Cyber-Physical System (CPS) that aims at transporting passengers from a floor to another one safely, while guaranteeing the highest comfort as possible. CPSs integrate digital cyber computations with physical processes~\cite{derler2011modeling,baheti2011cyber,alur2015principles}. The autonomy of such systems has considerably increased in the last few years, where most of their functionality is driven by software. As other CPSs, elevators also operate in a continuous dynamic environment, and because of this, they need to deal with unforeseen situations, due to the high uncertainty at which they are exposed to~\cite{han2022elevator,zhang2016understanding,menghi2019generating,zhang2019uncertainty,ali2015u}. 
For instance, such uncertainties could be due to the interaction of the system with humans~\cite{zhang2016understanding}. When such unforeseen situations arise, the CPS may often fail~\cite{han2022elevator,han2022uncertainty}. Providing an isolated failure-inducing test input to the CPS developer is of high benefit for two main reasons. Firstly, in the context of CPSs, such test inputs are usually based on operational data, which are long in time. Reproducing the situation using simulation-based testing may take a long time, as CPSs are usually compute-intensive~\cite{menghi2020approximation}. Secondly, with the minimal test input, the engineer can diagnose easier under which circumstances the CPS fails. Hence, reducing the failure-inducing test inputs helps (1) reducing the time it takes a test case to execute and (2) reducing debugging complexity by removing irrelevant elements to trigger the failure. With this, proposing a patch, will be easier and more agile. In our case study system, 55\% of such patches can be resolved through changing configuration parameters of the software~\cite{valle2023automated}. 


Delta debugging~\cite{hildebrandt2000simplifying,zeller2002simplifying} is an efficient and effective algorithm for isolating failure-inducing test inputs. In this paper we propose a novel extension of the delta debugging algorithm for isolating failure-inducing test inputs of CPSs. The core idea of this adaption is simple but effective. Although CPSs work in a continuous dynamic environment, such environment may eventually become static along the test execution. Our extended version of the delta debugging algorithm monitors the environment of the CPS to search for such static situations throughout the execution of a test case. Such static situations are subsequently used by the delta debugging algorithm to allow for both (1) a faster reduction time of the failure-inducing algorithm and (2) a more isolated failure-inducing test input. We adapt such approach to isolate the failure-inducing inputs of an industrial CPS, i.e., the dispatching algorithm of elevators. The evaluation is carried out with real unforeseen situations that appeared in operation with such system. 

The key contributions of this work can be summarized as follows:

\noindent \textbf{Technique:} We propose an adaption of the delta-debugging algorithm, named ``environment-wise delta debugging''. This algorithm extends the traditional delta debugging algorithm and is specifically designed for isolating failure-inducing test inputs of CPSs. Two variants are proposed, one of which reduces failure-inducing test inputs based on the order of events whereas the other based on the time that occur such events. 

\noindent \textbf{Application:} We adapt the algorithm to the context of elevator dispatching algorithms, integrating it with an industrial case study provided by Orona, one of the largest European elevator companies.

\noindent \textbf{Evaluation:} We evaluate the approach with this industrial case study, using real passenger data from operation, and where isolating failure-inducing inputs was paramount. The results suggest that our approach reduces the failure-inducing test inputs between 1.3 to 1.8 times faster than the traditional delta debugging algorithm, while producing a larger reduction in the test inputs. Moreover, we interview domain experts to qualitatively assess their opinion of the proposed techniques. 

\noindent \textbf{Lessons learned: }As a last contribution, we grasp a set of lessons learned from the application of the delta-debugging algorithm in an industrial context.

%% file: casestudy.tex
\begin{figure}[ht]
\centering
\includegraphics[width=0.45\textwidth]{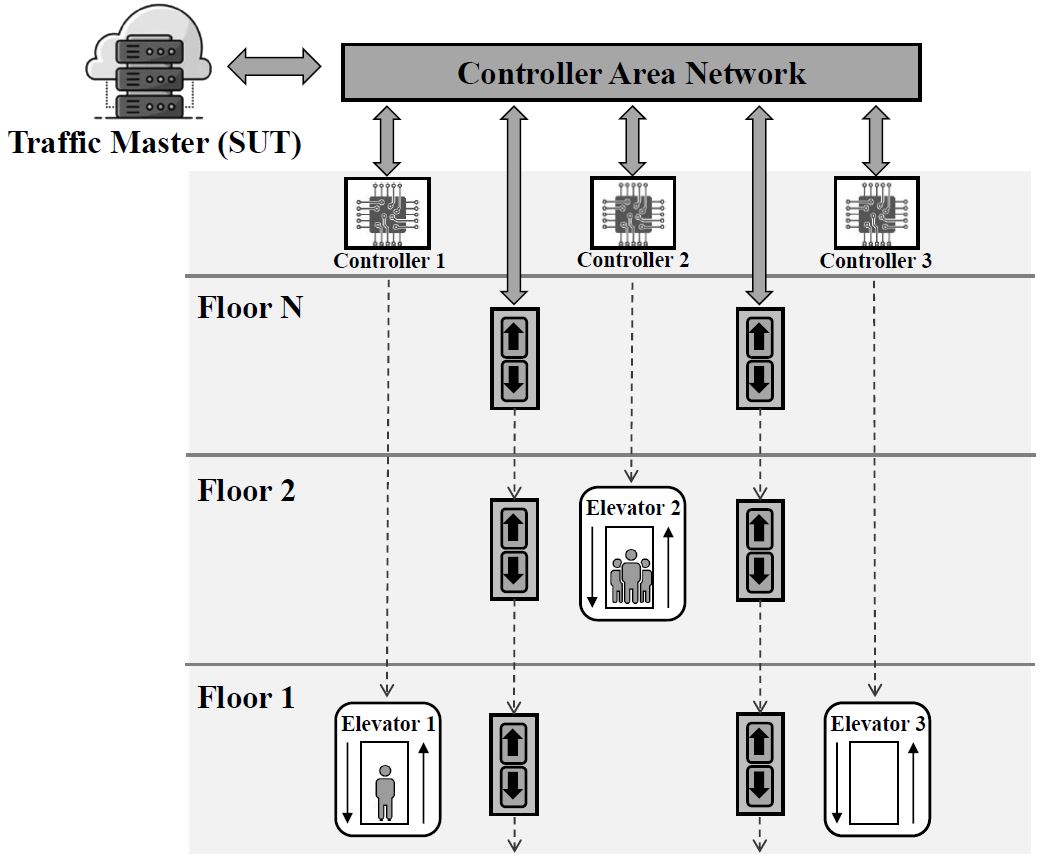}
\caption{Overview of our industrial case study}
\label{fig:casestudy}
\end{figure}

\textbf{Overview of the CPS:} Figure \ref{fig:casestudy} shows an overview of our industrial case study provided by Orona. A system of elevators is a complex CPS where different computational units, communication protocols, and mechanical and electrical components interact among them to transport passengers from a floor to another. Every time a passenger arrives at a floor, she/he makes a call through a button. This call is communicated to the traffic master through a Controller Area Network (CAN) bus. The traffic master is in charge of determining, for each call, which will be the elevator attending it. This is based on different criteria, such as, aiming to reduce the passengers' waiting times or the energy consumption. When the traffic master decides which elevator should be assigned to a call, this is communicated through the CAN bus to the controller in charge of moving the elevator.

\textbf{The System Under Test:} The System Under Test (SUT) in our study is the traffic dispatching algorithm, which is an important module of the traffic master. The traffic dispatching algorithm takes information of the environment, such as, position of each elevator or the number of passengers that each elevator has. Based on it, the algorithm returns the best solution to assign an elevator to each call. Naturally, this algorithm deals with much uncertainty (e.g., number of passengers behind a call, destination of the passenger, the weight of the passenger). Furthermore, in occasions, the system needs to deal with unforeseen situations for which the dispatching algorithm is not prepared or configured~\cite{han2022elevator,han2022uncertainty}. When such situations arise, it is of high benefit to isolate the situation as much as possible to propose a patch.

\textbf{Test execution platform:} For testing the traffic dispatching algorithm, different simulation test levels are employed~\cite{ayerdi2020towards}. The first level refers to the Software-in-the-Loop (SiL) test level, which is the level at which we have integrated the different versions of the delta debugging algorithm. In such level, the traffic dispatching algorithm is integrated with a commercial simulation tool named Elevate\footnote{https://elevate.helpdocsonline.com/home\label{elevate}}. Elevate takes as inputs two files: (1) the installation of the elevators (encompassing data like number of lifts, speed of each of them, number of floors, etc.) and (2) the test input, which involves a set of passengers traveling through the building. With these two files, Elevate simulates all the physical components of a system of elevators (e.g., engines, speed of the elevators) and provides as output a file with several information (e.g., the time that each passenger had to wait, which elevator attended each of the passengers, consumed energy). This file is parsed by an oracle to raise a test verdict. Our failure isolation algorithm is integrated with this test level. The following two test levels refer to the Hardware-in-the-Loop (HiL) test level and the level of Operation.\footnote{Not explained in this paper, but further details can be found in~\cite{ayerdi2020towards}} 

\textbf{Importance of isolating failure-inducing test inputs:} The test inputs for testing the traffic dispatching algorithm aim at mimicking realistic passenger data. Such data is often based on operational data, but synthetic data also mimic passengers traveling in a floor throughout an entire day~\cite{barney2015elevator}. Such test inputs involve thousands of passengers. However, the failure could be caused only by the interaction of a few of them. Obviously, debugging a failure is easier when only a few passengers exist than when many of them do. Furthermore, it is well-known that simulation-based testing is an expensive technique~\cite{menghi2020approximation,abdessalem2020automated,abdessalem2018testing,humeniuk2022search,arrieta2017employing,arrieta2016test,arrieta2019pareto,arrieta2019search,arrieta2023some}. Therefore, besides complexity, reducing the number of passengers in a test input helps decrease the test execution time.


%% file: approach.tex
We now explain our approach to reduce failure-inducing test inputs applied to the context of traffic dispatching algorithms.



\subsection{Formalization}\label{sec:Formalization}

A test input in the context of our system is a set of passengers traveling in a building through elevators. Let $TI$ be a test input composed on $np$ number of passengers, i.e., $TI = \{p_1, p_2, ..., p_{np}\}$. Each passengers has the following set of attributes: \begin{itemize}
    \item \textbf{Arrival time (at):} Time at which the passenger arrived to the floor and pushed the elevator button.
    \item \textbf{Arrival floor (af):} Floor at which the passenger arrived.
    \item \textbf{Destination floor (df):} Floor at which the passenger is traveling.
    \item \textbf{Mass (m):} Weight of the passenger. 
    \item \textbf{Capacity factor (cf):} The capacity of the elevator for which a passenger considers the elevator to be full. Based on studies and recommendations, the default value of this attribute is set to 80\%.
    \item \textbf{Entering time (ent):} Time the passenger takes to enter the elevator, by default 1.2 seconds.
    \item \textbf{Exiting time (ext):} Time the passenger takes to exit the elevator, by default, 1.2 seconds.
\end{itemize}

Let $p_i.attribute$ be the specific attribute of the passenger $i$ in $TI$. For instance, $p_2.at$ refers to the time at which the second passenger in $TI$ arrived.

\subsection{Running Example}\label{sec:runningExample}

As running example (Table \ref{tab:trafficProfile}) let us consider a test input (TI) of ten different passengers ($p_1, p_2, ..., p_{10}$), traveling in a building of 10 floors and three elevators. Each of these passengers has the aforementioned attributes. After executing such TI, Elevate returns a file with the result of the simulation. Table \ref{tab:schedule} shows an excerpt of such information, and includes, for each passenger (1) all the attributes in the TI, (2) which was the elevator taken by the passenger, (3) time at which the elevator arrived, (4) time at which the passenger reached the destination, (5) the waiting time (wt) of the passenger and (6) the transit time (tt) of the passenger (i.e., how much time the passenger spent inside an elevator). All this information is used by a test oracle that classifies a test execution either as a \textit{Pass} or as a \textit{Fail}. For instance, one can consider that a high waiting time of a passenger might be due to some sub-optimal assignment of an elevator to that call because of a bug or a misconfiguration.

\vspace{-0.25cm}

\begin{table}[htbp]
\scriptsize
 \centering
 
 \caption{\label{tab:trafficProfile} Example of a test input with 10 passengers 
 }
  \renewcommand\tabcolsep{2pt}
 \begin{tabular}{lccccccc}
  \toprule
  \textbf{ID}&\textbf{at} & \textbf{af} & \textbf{df}   & \textbf{m}  & \textbf{cf} & \textbf{ent}  & \textbf{ext}\\
  \midrule
 
$p_1$ & 11945 (3:19:05) & 1 & 8 & 75 & 80 & 1.2 & 1.2 \\
$p_2$ & 11985 (3:19:35) & 10 & 4 & 75 & 80 & 1.2 & 1.2 \\
$p_3$ & 12060 (3:21:00) & 3 & 1 & 75 & 80 & 1.2 & 1.2 \\
$p_4$ & 12080 (3:21:20) & 3 & 1 & 75 & 80 & 1.2 & 1.2 \\
$p_5$ & 12115 (3:21:55) & 1 & 8 & 75 & 80 & 1.2 & 1.2 \\
$p_6$ & 12130 (3:22:10) & 4 & 8 & 75 & 80 & 1.2 & 1.2 \\
$p_7$ & 12210 (3:23:30) & 3 & 1 & 75 & 80 & 1.2 & 1.2 \\
$p_8$ & 12365 (3:26:05) & 7 & 4 & 75 & 80 & 1.2 & 1.2 \\
$p_9$ & 12367 (3:26:07) & 7 & 4 & 75 & 80 & 1.2 & 1.2 \\
$p_{10}$ & 12375 (3:26:15) & 7 & 4 & 75 & 80 & 1.2 & 1.2 \\
  \bottomrule
 \end{tabular}\\
\end{table}

\vspace{-0.45cm}

\begin{table}[htbp]
\tiny
 \centering
 \caption{\label{tab:schedule} Partial simulation results corresponding to Table \ref{tab:trafficProfile}. \textit{WT} and \textit{TT} are in seconds. The \textit{ID} column is added to distinguish different passengers.}
  \renewcommand\tabcolsep{3pt}
 \begin{tabular}{lccccc}
  \toprule
  \textbf{ID}  & \textbf{Elevator Used} & \textbf{Time Elevator Arrived}   & \textbf{Time Reached Destination}  & \textbf{WT} & \textbf{TT} \\
  \midrule
$p_1$ &  2 & 3:19:55 & 3:20:31 & 48.4 & 35.7 \\
$p_2$ & 3 & 3:20:25 & 3:20:45 & 50.8 & 19.9 \\
$p_3$ & 1 & 3:21:20 & 3:21:45 & 19.6 & 25.3 \\
$p_4$ &  1 & 3:23:00 & 3:23:25 & 100.6 & 25.3 \\
$p_5$ &  1 & 3:25:20 & 3:25:55 & 205.2 & 35.3 \\
$p_6$ & 3 & 3:24:20 & 3:24:51 & 130.4 & 31.5 \\
$p_7$ &  1 & 3:23:50 & 3:24:15 & 19.6 & 25.3 \\
$p_8$ & 2 & 3:26:10 & 3:26:40 & 4.9 & 30.1 \\
$p_9$ & 2 & 3:26:10 & 3:26:40 & 8.9 & 30.1 \\
$p_{10}$ & 2 & 3:26:30 & 3:26:40 & 15 & 10.1 \\
  \bottomrule
 \end{tabular}\\
\footnotesize{The \textit{Passenger Attributes} are the seven attributes in Table \ref{tab:trafficProfile} corresponding to the specific passenger \textit{ID}. }
\end{table}

\vspace{-0.25cm}

\subsection{Adaption of the Delta Debugging Algorithm}

We now explain how we adapted the original delta debugging algorithm to our context. To this end, two versions of the algorithm have been implemented, named as (1) time-based delta debugging and (2) event-based delta debugging. The former refers to isolating the test inputs by considering the time at which the passengers arrived to the floor and made a call. The latter refers to isolating the test inputs by considering the order in which the passengers arrived to the floor. 



\subsubsection{Time Based Delta Debugging Algorithm (DD\textsubscript{TB})}\label{sec:TDD}
Algorithm \ref{alg:TimeDeltaDebugging} describes the adapted version of the Delta Debugging algorithm based on Time. As input, it receives the System Under Test (SUT), the initial failure-inducing Test Input ($TI$) and its Failing Time ($FT$), which indicates the simulation time at which the oracle detects the failure. As output it provides, $TI'$, which corresponds to the minimal failure-inducing Test Input for $TI$. In short, the algorithm returns a test input composed by different passengers by considering the time at which each passenger arrives. This is obtained by following the next procedures:

First, the algorithm splits the test input in order this to stop once the failure is triggered (Line 1, Algorithm \ref{alg:TimeDeltaDebugging}). This is performed because if the SUT fails at certain point, there is no need to carry out executing the remaining test. Therefore, test execution time is saved, making our algorithm more efficient. For instance, given the example in Section \ref{sec:runningExample}, suppose a test is considered \textit{fail} if the $wt$ of a passenger exceeds 200 seconds. For simplicity reasons, let us assume this is the only requirement we are targeting, although our approach can handle multiple. Our oracle will return a failure at time \textit{3:25:20}, when Elevator 1 attends passenger $p_5$, after this one needing to wait 205.2 seconds. In addition, in our case study, it is important to consider the next passengers. For instance, in this case, the time for attending $p_5$ could have been due to Elevator 1 stopping at floor 3 to attend $p_6$ and at floor 4 to attend $p_7$, although in the end, this passenger uses another elevator. Therefore, the split function that we implemented includes all passengers previous to the conflicting passenger (i.e., $p_5$ in this case) and those passengers after the conflicting passengers that may have influenced the allocation of an elevator to the conflicting passenger. Therefore, in that step, the algorithm will provide $TI'$ with passengers $p_1$ to $p_7$. $TI'$ in the algorithm represents the minimal failure-inducing input confirmed so far. The failure will be triggered without considering $p_8$ to $p_{10}$ because, by the time arriving those passengers, the test has already failed. We represent passengers from $TI'$ as $p'_1$ to $p'_{npr}$, $npr$ being the total number of passengers in $TI'$, and $npr \leq np$.

The following steps consist in reducing $TI$ by removing passengers that are before the conflicting passenger. Lines 2 - 4 in Algorithm \ref{alg:Time_Split_Minimizing} provide the first reduction. This process is carried out by invoking the function \textit{\textsc{SplitMinTime}}, which takes as inputs (i) $TI'$, (ii) the simulation time of $TI'$ ($simTime$) and (iii) the iteration of the algorithm ($it$). Because the employed simulator does not begin the simulation from second 0, but from the time at which the first passenger arrives, the simulation time is required to be obtained (Line 2). 
In this first iteration, the function returns the passengers that arrive after $p_1.at$ + $simTime$. Therefore, for the running example in Section \ref{sec:runningExample}, the function returns passengers $p_4$ to $p_7$. 


After this, the algorithm enters in a while loop (Lines 5-14) that tries to minimize the failure inducing test input as much as possible. To this end, it first executes the test in $TI_{NEW}$ (Line 6). If the test returns a failure, the minimization procedure can continue. The test input in $TI_{NEW}$ is assigned to $TI'$, and the minimization routine is invoked again by means of the \textit{\textsc{SplitMinTime}} function. If the test has passed, it means that the test input was minimized too much. Therefore, the test input requires to be enlarged (i.e., more passengers are required to reproduce the failure). This is carried out by invoking the routine in Algorithm \ref{alg:Time_Split_Maximizing}, which adds a number of passengers between the number of passengers in $TI_{NEW}$ and $TI'$.
 This procedure returns the maximized test input in $TI_{NEW}$, which is tested in Line 6 of Algorithm \ref{alg:TimeDeltaDebugging}. This process is repeated until the number of passengers in $TI'$ and $TI_{NEW}$ are the same. When this condition is given, the delta debugging algorithm returns the minimal failure-inducing test input.

\begin{algorithm}[ht]
\caption{Time-based Delta Debugging Algorithm}\label{alg:TimeDeltaDebugging}
    \KwIn{SUT \tcp*{\small System Under Test} \\
    \nonl FT \tcp*{\small Failing time}\\
    \nonl TI = $\{p_1, p_2,..., p_{np}\}$   \tcp*{\small Initial failure inducing test input} \\}
    \KwOut{TI'= $\{p'_1, p'_2,..., p'_{npr}\}$ \tcp*{\small Minimized failure inducing test input} \\}
    TI' $\gets$ \textsc{Split}(TI,FT); \\ 
    simTime $\gets$ $p'_{npr}.at - p'_1.at$;\\
    it $\gets$ 1;\\
    TI\textsubscript{NEW} $\gets$ \textsc{SplitMinTime}(TI', simTime, it);\\
    \While {$TI'.np$ $\neq$ $TI\textsubscript{NEW}.np$}{
       
        Verdict $\gets$ \textsc{ExecuteTest}(TI\textsubscript{NEW}, SUT);\\
        it $\gets$ it+1;\\
        \eIf{Verdict == Failure}{
            TI' $\gets$ TI\textsubscript{NEW};\\
            TI\textsubscript{NEW}  $\gets$ \textsc{SplitMinTime}(TI\textsubscript{NEW}, simTime, it);\\
        }{
            TI\textsubscript{NEW} $\gets$ \textsc{SplitMaxTime}($TI\textsubscript{NEW}$,TI', simTime, it);\\
        }
        
    }
    
\end{algorithm}


\begin{algorithm}[ht]
\caption{\textsc{SplitMinTime}: Time-based split Minimizing}\label{alg:Time_Split_Minimizing}
    \KwIn{
    \nonl simTime \tcp*{\small Total simulation time} \\
    \nonl it \tcp*{\small Number of iterations}\\
    \nonl TI = \{$p_1$, $p_2$, ..., $p_{np}$\} \tcp*{\small Minimized test input previously selected}\\}
    \KwOut{TI\textsubscript{NEW} \tcp*{\small Minimized test input}\\}
    splitTime $\gets$ $p_1.at + simTime/ 2^{it}$; \\
    
    \For{$i\gets1$ \KwTo $TI.np$}{
        \If{$p_i.at$ $>$ splitTime }{
            $TI\textsubscript{NEW}$ $\leftarrow$ $TI\textsubscript{NEW} \cup p_i$;\\
        }
    }
\end{algorithm}



\begin{algorithm}[ht]
\caption{\textsc{SplitMaxTime}: Time-based split Maximizing}\label{alg:Time_Split_Maximizing}
    \KwIn{
     simTime \tcp*{\small Total simulation time}\\
    \nonl it \tcp*{\small Number of iterations} \\
    \nonl TI\textsubscript{NEW} =$\{p'_1, p'_2,..., p'_{npr}\}$ \tcp*{\small Minimized failure inducing test input}   \\
    \nonl TI = $\{p_1, p_2,..., p_{np}\}$ \tcp*{\small Minimized test input previously selected} \\}
    \KwOut{TI' \tcp*{\small Maximized test input} \\}
    splitTime $\gets$ $p'_1.at - simTime/2^{it}$; \\
    \For{$i\gets0$ \KwTo $TI.np$}{
        \If{$p_i.at$ $>$ splitTime }{
            $TI'$ $\leftarrow$ $TI' \cup p_i$\\
        }
    }
\end{algorithm}


\subsubsection{Event-Based Delta Debugging Algorithm (DD\textsubscript{EB})}

Algorithm \ref{alg:PassengerDeltaDebugging} describes the adapted version of the Delta Debugging algorithm based on events. Unlike the time-based delta debugging, this version reduces the failure-inducing test inputs by considering the order at which passengers arrive, instead of their arrival time. 

As in the time-based approach, the event-based delta debugging starts by minimizing the failure-inducing test input in order the test execution to stop once the failure is triggered (Line 1, Algorithm \ref{alg:PassengerDeltaDebugging}). The same splitting routine as the previous algorithm is used. Therefore, from the example in Section \ref{sec:runningExample}, our oracle will return a test input with passengers $p_1$ to $p_7$. After, the algorithm starts isolating the failure by first splitting the test input by reducing the number of passengers by half of it (Lines 2-3, Algorithm \ref{alg:PassengerDeltaDebugging}). As there are 7 passengers in $TI'$, passengers $p_4$ to $p_7$ will be returned by the algorithm. 

As in the previous case, the algorithm enters in a while loop to minimize the failure-inducing test input as much as possible (Lines 4-13, Algorithm \ref{alg:PassengerDeltaDebugging}). First, we obtain the verdict of the newly reduced test input (Line 5, Algorithm \ref{alg:PassengerDeltaDebugging}). If the test returns a failure, the algorithm tries to reduce even more the failure-inducing input (Lines 8-9, Algorithm \ref{alg:PassengerDeltaDebugging}). Conversely, if the test returns a \textit{pass}, it means that the test input has been reduced too much, therefore, requiring more passengers (Line 11, Algorithm \ref{alg:PassengerDeltaDebugging}). For instance, if $TI_{NEW}$ has passengers $p_4$ to $p_7$, and $TI'$ has passengers $p_1$ to $p_7$, a new test input with passengers $p_2$ to $p_7$ will be created, and tested again. This process is iteratively repeated until the number of passengers in test input $TI'$ and $TI_{NEW}$ are the same, which means that it is not possible to further reduce the test input.


\begin{algorithm}[ht]
\caption{Event-based Delta Debugging Algorithm}\label{alg:PassengerDeltaDebugging}
    \KwIn{SUT \tcp*{\small System Under Test}  \\
        \nonl FT \tcp*{\small Failing Time} \\
        \nonl TI $\{p_1, p_2,..., p_{np}\}$ \tcp*{\small Initial failure inducing test input}  \\}
    \KwOut{TI' $\{p'_1, p'_2,..., p'_{npr}\}$ \tcp*{\small Minimized failure inducing test input}\\}
    TI' $\gets$ \textsc{Split}(TI,FT); \\ 
    splitSize $\gets$ $\lceil$$TI'.np$/2$\rceil$;\\
    TI\textsubscript{NEW}  $\gets$  \textsc{SplitMinEvent}(TI', splitSize);\\
    
    \While {$TI'.np$ $\neq$ $TI\textsubscript{NEW}.np$}{
        Verdict $\gets$ \textsc{ExecuteTest}(TI\textsubscript{NEW}, SUT);\\
        splitSize $\gets$ $\lceil$splitSize/2$\rceil$;\\
        \eIf{Verdict  == Failure}{
            TI' $\gets$ TI\textsubscript{NEW};\\
            TI\textsubscript{NEW}   $\gets$  \textsc{SplitMinEvent}(TI\textsubscript{NEW}, splitSize);\\
        }{
            TI\textsubscript{NEW}  $\gets$  \textsc{SplitMaxEvent}($TI\textsubscript{NEW}.np$,TI', splitSize);\\
        }
    }
\end{algorithm}


\begin{algorithm}[ht]
\caption{\textsc{SplitMinEvent}: Event-based split Minimizing}\label{alg:Passenger_Split_Minimizing}
    \KwIn{
    \nonl splitSize \tcp*{\small Point to split}    \\
    TI=$\{p_1, p_2,..., p_{np}\}$ \tcp*{\small Minimized test input previously selected}\\}
    \KwOut{TI\textsubscript{NEW} \tcp*{\small Minimized test input} \\}
    
    \For{$i\gets splitSize$ \KwTo $TI.np$}{
        
        $TI\textsubscript{NEW}$ $\leftarrow$ $TI\textsubscript{NEW} \cup p_i$;\\
        
    }
\end{algorithm}


\begin{algorithm}[ht]
\caption{\textsc{SplitMaxEvent}: Event-based split Maximizing}\label{alg:Passenger_Split_Maximizing}
    \KwIn{ sizeTI \tcp*{\small \# of selected passengers} \\ 
    \nonl splitSize \tcp*{\small Point to split} \\
    \nonl TI = $\{p_1, p_2,..., p_{np}\}$ \tcp*{\small Minimized test input} \\}
    \KwOut{TI\textsubscript{NEW} \tcp*{\small Maximized test input} \\}
    toSplit  $\gets$  $TI.np$-(sizeTI + splitSize); \\
    \For{$i\gets toSplit$ \KwTo $TI.np$}{
        \
        $TI\textsubscript{NEW}$ $\leftarrow$ $TI\textsubscript{NEW} \cup p_i$;\\

    }
\end{algorithm}

\subsection{Environment-wise Delta Debugging Algorithm (EWDD)}
\label{sec:EWDD}

We now explain the proposed environment-wise delta debugging algorithm (EWDD), shown in Algorithm \ref{alg:DeltaDebugging_EnvironmentAware}. First, we obtain the Environmental States (ES) of the CPS while this has been tested (Line 1). With this information, we search for, what we call, \textit{``static states''} of the system. Such static states refer to stable states of the CPS and its environment. For instance, in our industrial case study, we consider the CPS is in a static state when all elevators are stopped, with no passengers inside the cabins and the doors of all elevators closed. When executing a test, several static states can be found throughout the simulation. After inspecting the environment states, we initialize variable $i$ (Line 2) to the total number of static states found, and we split the original failure inducing test input (i.e., $TI$) (in Line 3). As a next step, the algorithm enters into a do-while loop (Lines 4-10). 

In this loop, the algorithm first gets the last static state found before the system failed (Line 5). This static state is an object that contains several information (e.g., the time the static state started and finished, position of elevators). With this static state, the CPS and its environment is configured to execute the test (Line 6). It is assumed that the algorithm can somehow configure the CPS and its environment to start the simulation from such a static state. For instance, in our industrial case study, it is possible to prepare the simulation environment to execute a test with each elevator in certain position. After preparing the CPS and its environment, we reduce the test input by removing all the passengers before the static situation was given, obtaining $TI_{NEW}$ (Line 7). We execute the test with $TI_{NEW}$ (Line 8) and reduce the variable $i$ by 1, such that the previous static state is considered in case the algorithm was unable to reproduce the failure.

Once the failure is reproduced, the original delta debugging algorithm is executed. With this algorithm, we conjecture that searching for static states of the CPS can help reduce the failure-inducing test input. This is because the CPS is in a stable position. Of course, cumulative variables might arise, which may lead the CPS to take other decisions. This could therefore result in not reproducing the failure. In such a case, the algorithm would find the previous static state. This approach may have two core limitations. Firstly, domain expertise is required to define the so-called static states. Secondly, a CPS might be very complex and might be data-intensive. In those cases, it might be computationally too expensive to search for static situations. 

\begin{algorithm}[ht]
\caption{Environment-wise Delta Debugging Algorithm}\label{alg:DeltaDebugging_EnvironmentAware}
      \KwIn{SUT \tcp*{\small System Under Test} \\
        \nonl FT \tcp*{\small Failing Time} \\
        \nonl SD \tcp*{\small Simulation Data}\\
        \nonl TI = $\{p_1, p_2,..., p_{np}\}$ \tcp*{\small Initial failure inducing test input} \\}
    \KwOut{TI' = $\{p'_1, p'_2,..., p'_{npr}\}$ \tcp*{\small Minimized failure inducing test input}\\}
    
    ES  $\gets$ \textsc{GetEnvironmentStatesUntilFailure}(SD,FT);\\
    i $\gets$ \textsc{NumberOfStaticStates}(ES); \\
    TI' $\gets$ split(TI,FT); \\ 
    \Do{Verdict == Pass}{
        staticState $\gets$ \textsc{GetState}(ES,i); \\
        \textsc{PrepareEnvironment}(staticState);\\
        $TI_{NEW}$ $\gets$ \textsc{SplitStaticState}(staticState,TI');\\
        Verdict $\gets$ \textsc{ExecuteTest}($TI_{NEW}$); \\
        i $\gets$ i-1; 
    }
    Execute Algorithm \ref{alg:TimeDeltaDebugging} or \ref{alg:PassengerDeltaDebugging} from Line 2 on
    
\end{algorithm}

\subsection{Test Oracles in Cyber-Physical Systems}

Our approach assumes the availability of a test oracle that provides a test verdict in the form of \textit{pass} or \textit{fail}. However, it is noteworthy that in the context of CPSs, test oracles are not usually Boolean, but also provide a quantitative measure of the severity of a failing requirement~\cite{abdessalem2020automated,arrieta2022automating,menghi2019generating,abdessalem2018testing,humeniuk2022search}. For a failing test input ($ti$), it can expose one or more failure(s) whose severity is $O(ti),r_j$ $\in$ [-1; 0], where $r_j$ is the violated $j$-th requirement. The lower this value, the higher the severity of the failure. For instance, let us assume that we are testing the maximum waiting time (MWT) of passengers, where it should be below 70 seconds. However, the severity level of a test case with MWT of 400 seconds is higher than the severity of another test case with 100 seconds.

When aiming at reducing failure-inducing test inputs, it is important to maintain the level of severity as high as the original failure-inducing test input. That is, following with the above example, if a test input is considered \textit{fail} if the maximum waiting time is 70 seconds, but the failure-inducing input deviates the waiting time to 400 seconds, it is more interesting to the engineer to provide a test input which is closer to the 400 seconds rather than to the 70 seconds. This is due to the severity of the violated requirement.

For this reason, we provide certain thresholds to the test oracle based on the original failure-inducing test input. In our industrial case study, our oracle checks different metrics, such as the waiting time each passenger has. 
 The threshold allows for specifying a minimum severity degradation over the original test input. For the example above, if for the original test input the passenger waiting the longest time had to wait 300 seconds, and we specify a threshold of 5\%, the test oracle will return a \textit{fail} only if the waiting time of such passenger is not reduced more than a 5\% (i.e., if the passenger waits less than 285, the oracle returns a pass). This way, the failure-inducing test input maintains a minimum severity level obtained by the original test input. 

Note that our approach is also applicable with multiple requirements. Each requirement is modeled with its own oracle providing its own severity level, and the threshold can be the same or different for each of these oracles. If any of these fail, the test is considered to be failed. 


%% file: evaluation.tex
In this section, we empirically evaluate the proposed approach by means of an industrial case study. We aimed at answering the following four research questions (RQs):


\begin{itemize}
    \item \textit{RQ1 -- Is the problem of isolating failure-inducing test inputs non-trivial in this context?} 
    \item \textit{RQ2 -- Is the proposed environment-wise delta debugging more efficient and effective than the traditional one?} 
    \item \textit{RQ3 -- Is there any difference between the time-based and event-based techniques?}
    
    \item \textit{RQ4 -- How useful do domain experts find the produced failure-inducing test inputs?}
    
    
\end{itemize}

The first RQ is a sanity check that aims at studying whether isolating failure-inducing test inputs in this context is a non-trivial problem; to this end, we implemented a baseline algorithm (named as backward strategy) and compared our approach against it. The second RQ aims at comparing our proposed environment-wise delta debugging algorithm with the traditional one. With the third RQ we aimed at studying if there is any difference between the event-based and time-based strategies. Lastly, we aim at qualitatively analyzing the opinion of the developers regarding the proposed methods.

\subsection{Experimental Set-up}
\subsubsection{Baseline Algorithm} \label{sec:Baseline}

To answer the first RQ, we implemented a baseline algorithm. This baseline algorithm takes the conflicting passenger and those passengers after the conflicting one that may affect the failure-inducing input. After that, it includes one by one the remaining passengers. For instance, for the example in Section \ref{sec:runningExample}, $p_5$,$p_6$ and $p_7$ are considered as the beginning test input. After that, the algorithm starts a backward strategy where passengers are included one by one until the failure is reproduced (i.e., first, passenger $p_4$ is included, if failure not reproduced, $p_3$ is included, etc.). The baseline algorithm is shown in Algorithm \ref{alg:BAckward}.




\begin{algorithm}[ht]
\caption{Backward Isolation Algorithm (Baseline)}\label{alg:BAckward}
    \KwIn{SUT \tcp*{\small System Under Test} \\
        \nonl FT \tcp*{\small Failing Time} \\

        \nonl TI = $\{p_1, p_2,..., p_{np}\}$ \tcp*{\small Initial failure inducing test input} \\}
    \KwOut{TI\textsubscript{NEW} = $\{p'_1, p'_2,..., p'_{npr}\}$ \tcp*{\small Minimized failure inducing test input}\\}
    TI' $\gets$ \textsc{GetConflictPassengers(TI,FT)}; \\
    it $\gets$ \textsc{ConflictPassengerID(TI')};\\
    \Do{Verdict == Pass}{
        TI\textsubscript{NEW}  $\gets$ TI\textsubscript{NEW} $\cup$ $p_{it}$;\\
        Verdict $\gets$ \textsc{ExecuteTest}(TI\textsubscript{NEW}, SUT);\\
        it=it-1;\\
    }
\end{algorithm}



\subsubsection{System Under Test and Test Inputs Characteristics} 

We used Orona's Conventional Group Control (CGC) traffic dispatching algorithm~\cite{barney2015elevator}, which is the most well-known algorithm in the domain. Furthermore, we used a real installation to assess our approach. This installation was used because certain unforeseen situations were detected (e.g., abnormal passenger flow). Moreover, for those situations, Orona obtained operational data from the installation. This allowed us to assess our approach with real operational data as failure-inducing test inputs. In total, we used three failure-inducing test inputs, which were those obtained by Orona in the conflicting building. Table~\ref{table:characteristics} summarizes the key characteristics of the used passenger files. Such data includes information of the passengers traveling in a building through 16 hours.

\begin{table}[ht]
\centering
\caption{Characteristics of the executed passenger files}
\label{table:characteristics}
\resizebox{0.49\textwidth}{!}{ 
\begin{tabular}{rrrrr}
\hline
\multicolumn{1}{l}{Test} & \multicolumn{1}{l}{\# of passengers} & \multicolumn{1}{c}{\begin{tabular}[c]{@{}c@{}}\# of passengers \\ to reproduce the failure\end{tabular}} & \multicolumn{1}{l}{Execution Time (s)} & \multicolumn{1}{l}{Failing Time (s)} \\ \hline
Test 1 & 3,769 & 2,330 & 57,463 & 29,261 \\
Test 2 & 3,105 & 1,405 & 55,540 & 21,815 \\
Test 3 & 3,294 & 2,210 & 56,544 & 34,445 \\ \hline
\end{tabular}
}
\end{table}

\subsubsection{Execution Platform}

The simulator we used for executing the tests was Elevate version 8.19. The experiments were executed on a PC using a Windows 10 operating system, with a CPU Intel Core i5 7th generation, with a 16 Gb RAM. 









\subsubsection{Evaluation Metric and Statistical Tests for RQ1, RQ2 and RQ3}


We assessed our approach from two perspectives: (1) efficiency of the algorithms and (2) effectiveness of the algorithms. The former relates to how fast the algorithm is able to provide the minimal failure-inducing test input. The latter relates to the quality of the provided failure-inducing test input.

\textbf{Efficiency:} To assess the efficiency of the proposed approach, we measured the execution time required by the algorithms to return the failure-inducing test inputs. Since the execution time is stochastic, we ran each algorithm instance 20 times. To analyze the performance between pairs of algorithms (e.g., baseline with environment-wise delta debugging), we used different statistical tests. On the one hand, we employed the Mann-Whitney U-test to assess the statistical significance between two algorithms for the 20 different runs. This test was selected because the data was not normally distributed. We analyzed how the data was distributed by employing the Shapiro-Wilk test. The Mann-Whitney U-test provides a p-value that evaluates whether there was statistical significance (p-value$<$0.05) between two techniques. Besides, we used the Vargha and Delaney \^{A}$_{12}$ value to assess the effect size between two techniques. Unlike the Mann-Whitney U-test, the Vargha and Delaney \^{A}$_{12}$ value tells us which algorithm is better and it is very intuitive: in our context, it measures the probability that running Algorithm A will be \textit{faster} than running Algorithm B.

\textbf{Effectiveness:} Effectiveness is concerned with the quality of the test inputs returned by the different algorithms. We have measured the effectiveness from two different perspectives: (1) test execution time and (2) complexity of the test inputs. It is important to recall that the outcome of the algorithms are deterministic, and therefore, unlike for efficiency, for effectiveness it is not necessary to carry out any type of statistical test. Because the test execution time when testing CPSs is high, on the one hand, we assessed the effectiveness by employing the Test Input Reduction Ratio with respect to the failing time ($TIRR_{ft}$), proposed in our previous work \cite{valletowards}. This metric measures the test execution time reduction ratio obtained by the test input provided by the delta debugging algorithm until the failure is detected with respect to the time taken by the original test input to fail, calculated as follows: 

\begin{equation}
    TIRR_{ft}=1-\dfrac{tet_{fail}(TI')}{tet_{fail}(TI)}
\end{equation}

\noindent where $tet_{fail}(TI')$ stands for the test execution time needed to trigger a failure for the test input returned by the failure-inducing test input minimization algorithm (e.g., delta debugging) and $tet_{fail}(TI)$ stands for the test execution time needed to trigger a failure by the original test input. The higher this value, the higher the effectiveness of the algorithm.

On the other hand, we wanted to assess how complex the returned test inputs were. Complexity in our context relates to how many passengers the test input encompasses. It is basically impossible to isolate under what circumstances a system of elevators fails if the test inputs have thousands of passengers, as it happens in our original test inputs. To this end, for measuring complexity, we provide the final number of passengers of the test inputs returned by the algorithms. This metric provides an intuitive measure of the complexity of the returned test input. That is, the lower the number of passengers in the test input, the easier to diagnose under what situation the system is failing. 

The oracle was also configured to trigger a failure based on different severity levels. This allowed us to analyze the proposed algorithms under different situations (i.e., being strict with the severity level or being softer). We used a total of ten different threshold levels, ranging from 5\% to 20\%. Therefore, combining all these ten severity levels with three different failure-inducing test inputs, we had a total of 30 experimental scenarios.

\subsubsection{Qualitative Analysis of the Methods (RQ4)}

For RQ4, we qualitatively analyzed the produced failure-inducing test inputs by our approach with domain experts. We carried out semi-structured interviews with 4 domain experts working on the development of the traffic dispatching algorithms. Two of the domain experts had more than 10 years of experience with the system, another one had 5 years and the last one had 3 months. The interviews aimed at assessing the following \textit{three qualitative aspects}: 1) Whether the produced test inputs could help identify the root cause of the failure. 2) Whether the produced test inputs could help identify a patch. 3) Whether there are other advantages by the provided test inputs. We prepared one specific question for each of these aspects. As the interviews were semi-structured, we also asked follow-up questions based on the answers of the interviewed person.

\input{Tables/Mediak}

\subsection{Analysis of the Results and Discussion}

\subsubsection{RQ1 -- Sanity Check} The first RQ is employed as a sanity check to analyze whether isolating failure-inducing inputs is a non-trivial problem in our application context. To this end, we implemented a simple baseline algorithm (described in Section \ref{sec:Baseline}) and compared it against the traditional delta debugging algorithms and our proposed environment-wise delta debugging algorithm. Table \ref{table:TimeResults} shows the mean, median and standard deviation values for each of the techniques. As can be observed, when considering the algorithms' running times, the baseline technique was in the order of thousands seconds to isolate the failure-inducing test inputs, whereas the delta debugging techniques were in the order of hundreds. On the one hand, in the best of the cases, when considering the average values, the environment-wise delta debugging algorithm was 17.35 times faster than the baseline technique (i.e., for Test Input 1, $EWDD_{TB}$ algorithm and 5\% Threshold). On the other hand, in the worst of the cases, the traditional delta debugging was 4.31 times faster than the baseline (i.e., for Test 3, $DD_{EB}$ algorithm and 13\% Threshold). All these results were further corroborated by means of statistical tests. In all the cases, there was statistical significance (i.e., p-value $\leq$ 0.05), with the Vargha and Delaney \^{A}$_{12}$ being 1 in all the cases (i.e., probability of 100\% in favor of the delta debugging algorithms).

In terms of effectiveness, it can be observed that the reduction of the baseline technique was, in many of the cases, larger than the reduction provided by the delta debugging algorithms. Specifically, this happened in the case of the first failure-inducing test input. However, for test inputs 2 and 3, at least one of the environment wise delta debugging algorithms provided a larger reduction. This is mainly because our proposed approach is able to change the position of the elevators based on the observed static situations, whereas the backward approach does not have this possibility. 

The reason for providing a larger reduction in the first test input could have been the large severity of the failure-inducing test input in such a case. While it is true that when considering test input 1 the reduction when employing the backward strategy is larger than when employing our approach, it must be highlighted that the execution time is much higher too. Specifically, the average time for the first case (test input 1 and threshold of 5\%) was around 1.5 hours for the backward strategy. Conversely, the environment-wise algorithms could provide a solution within 5 minutes approximately. Therefore, the backward strategy seems not to scale for complex problems, answering the first RQ as follows:

\begin{tcolorbox}
\textbf{RQ1}: The problem of isolating failure-inducing test inputs is non-trivial in the context of elevator dispatching algorithms. This is mainly demonstrated due to the high execution cost of the backward strategy.
\end{tcolorbox}

\subsubsection{RQ2 -- Environment-wise vs Traditional Delta Debugging} 

This RQ aimed at comparing our proposed environment-wise delta debugging algorithm with the traditional one. 
For all the cases, there was statistical significance in favor of the environment-wise delta debugging algorithm, with very large effect sizes (i.e., all the \^{A}$_{12}$ values were above 0.99)\footnote{We omit adding the table for space reasons}. Based on our experiments, when considering the average running times, for the Test 1 and 19\% threshold, the environment wise delta-debugging was, on average, 1.31 times faster than the traditional delta debugging. This was the smallest difference that we found in all the experimental scenarios. As for the largest difference, in Test 1 for the threshold 10\%, the environment-wise algorithm was, on average, 1.83 times faster.


When considering the reduction ratio, we found that the environment wise delta debugging is significantly more effective than the traditional one. For instance, when considering the first failure-inducing test input at low threshold values (thresholds 5 to 11\%), the traditional delta debugging techniques provided test inputs that encompassed twice the number of passengers in the test inputs when compared with the environment-wise techniques. Within the remaining approaches, we found that, at least, one of the environment-wise delta debugging techniques provided a larger reduction when compared to the traditional ones. Thus, the  second RQ can be answered as follows:



\begin{tcolorbox}
\textbf{RQ2:} 
The environment-wise delta debugging is both more efficient and more effective than the traditional delta debugging algorithm. 
\end{tcolorbox}

\input{Tables/RQ3Table}

\subsubsection{RQ3 -- Event-based vs Time-based Delta Debugging} 

This RQ aims at studying whether there is difference when employing a time-based or an event-based strategy combined with the delta debugging algorithm. Table \ref{table:RQ3} summarizes the statistical tests results when measuring the algorithms' efficiency. As for the algorithms' running times, when considering the environment-wise delta debugging approach, all the \^{A}$_{12}$ values were in favor of the time-based strategy. Furthermore, in 28 out of 30 cases, there was statistical significance. Therefore, it can be observed that the time-based strategy outperformed the event-based when considering the environment-wise delta debugging approach. Nevertheless, when considering the running times, it could be observed that, on average, the maximum speed-up for the time-based strategy is 1.2 times faster than the event-based, which is not a high difference. 

When comparing the traditional delta debugging algorithms, there was no statistical significance in 8 out of the 30 cases. For the remaining, the statistical significance was in favor of the time-based strategy in 10 out of 22 cases. Interestingly, all these cases were in the third test input. The results were in favor of the event-based strategy for the remaining 12 cases. These results suggest that for the case of the traditional delta debugging algorithm, the efficiency of the strategies highly depends on the composition of the failure-inducing test input. When considering the average running times, it could be observed that in those cases where the time-based strategy is better, at most, it is 1.2 times faster than the event-based strategy. On the other hand, the event-based strategy is at most 1.1 faster, when the event-based strategy outperforms the time-based strategy. 

In terms of the effectiveness, we found mixed answers. In the case of the environment-wise delta debugging algorithms, results were the same for the test inputs 1 and 3. In the case of the test input 2, the event-based strategy showed a larger reduction when the oracle thresholds were configured for 5 and 7\%. However, for the remaining thresholds, the reduction by the time-based strategy was much larger (i.e., 36 passengers in total for the time-based strategy, whereas 65 passengers for the event-based). When considering the traditional delta debugging, results were also different depending on the threshold values of the test inputs. In the cases of test inputs 1 and 2, the reductions were larger by the time-based strategy for lower thresholds. In terms of number of passengers, some differences were large in those cases (e.g., 1,068 passengers vs 742 passengers). In the case of the third test input, the event-based strategy outperformed the time-based strategy in terms of the test input reduction ratio. However, the difference was not large (i.e., not more than 4 passengers). Thus, we can answer the third RQ as follows:

\begin{tcolorbox}
\textbf{RQ3:} We did not find clear differences between the event and time-based strategies. While there are differences in certain scenarios, there is no a clear winner in terms of an event-based vs time-based strategies.



\end{tcolorbox}









\subsubsection{RQ4 -- Qualitative Assessment} 

Within the fourth RQ, we aimed at qualitatively assessing the test inputs provided by the different instances of the delta debugging algorithm as well as the baseline technique. We carried out semi-structured interviews to four domain experts of different seniority levels. Within the interview, we provided the engineers with different test inputs. 

When assessing \textbf{\textit{whether the produced test inputs could help identify the root cause of the failures}}, all the four engineers agreed that a lower number of passengers in the test inputs helps identify the root cause of the problem much better. Indeed, for the test inputs 2 and 3, provided by the $EQDD_{EB}$ for the threshold levels of 9\% and higher (only in the case of the test input 2), the two most experienced engineers claimed to guess what the problem was by only seeing the test input and the test output. They also claimed that a test execution would only be required to confirm their guess. Within a follow-up question, both engineers claimed that the root cause starts to be more blurry to be guessed without a test execution when there are more than 60 to 70 passengers. On the other hand, the less experienced engineers claimed that they would not be confident enough to guess the root cause without executing the test input and checking in the user interface of the simulator the behavior of the elevators. When executing the test inputs 2 and 3, all engineers saw it clear which the problem was when executing the test inputs with any of the test inputs provided by the five algorithms within the established threshold of 5\%. Conversely, within test inputs 1 for the threshold of 11\% and below, none of them could clearly identify the root cause when using the traditional delta debugging, while with the test inputs provided by the event-wise and the backward techniques, they claimed the root cause could be easier guessed. 

When assessing \textbf{\textit{whether the produced test inputs could help identify a patch}}, the three most experienced engineers provided a positive answer. Indeed, the three of them proposed a change in the parameters of the dispatching algorithm. The three of them proposed two specific parameters (all of them the same, with the same values). The most experienced expert  also proposed additional alternatives, although she claimed that this would require further tests. Conversely, the less experienced engineer claimed to have low expertise yet to propose a patch.

Lastly, we assessed whether the domain experts found any other advantages \textbf{\textit{when using the test inputs}}. Two of them claimed that the technique could be eventually used to obtain critical parts of full-day passenger profiles in order to execute critical regions in their \textit{Hardware-in-the-Loop} environment. In such cases, the simulation is in real-time, and executing passenger profiles of 12 hours is not always viable. They all claimed that, in principle, the execution time in the PC simulator is not an issue, although the third engineer claimed that the technique could also be used to accelerate configuration optimization processes. 

\begin{tcolorbox}
\textbf{RQ4:} The higher the reduction of the test input, the easier engineers see which the root cause is, as well as, which could be a potential patch.
\end{tcolorbox}

\subsection{Threats to Validity}

Our evaluation was subject to some threats to validity. We now explain how we tried to mitigate them:

An \textbf{\textit{internal validity}} threat could be related to the parameters of our algorithms. The only configurable parameter was the threshold given to the oracle to raise a verdict as fail or as pass. To reduce such threat we employed 10 different threshold values. 

An \textbf{\textit{external validity}} threat in our evaluation relates to the generalizability of the results. We distinguished two dimensions for generalizability. On the one hand, obtaining the same level of benefits observed in our evaluation. On the other hand, applicability of our approach beyond our case study system. Regarding the former, additional case studies are necessary to further validate our approach, although our case study was performed in a representative industrial setting. Furthermore, we employed three failure-inducing test inputs, which were obtained from a real installation in operation. Regarding the latter, Section~\ref{sec:EWDD} provides a set of assumptions of how to apply the environment-wise delta debugging: (1) need to define a stable state of the CPS and (2) possibility to configure the CPS and its environment to start the simulation from such static state. For instance, Ardupilot\footnote{\url{ https://ardupilot.org/copter/docs/common-mavlink-mission-command-messages-mav_cmd.html}} has this option for the simulation of Unmanned Aerial Vehicles. Based on our experience of, both of these assumptions are generally common in CPSs simulators. Another \textbf{\textit{external validity}} threat refers to the use of humans to answer RQ4. To mitigate such threats, the involved experts were those developing dispatching algorithms at Orona. 

A \textbf{\textit{conclusion validity}} threat relates to the non-deterministic nature of the running times of our algorithms. We mitigated this threat by running each algorithm 20 times and employing statistical tests to analyze the results.

%% file: Tables/Mediak.tex
\begin{table*}[ht]
\centering
\caption{Summary results of the execution time for the performed experiment. For the test execution time, we provide the Median ($m$), Mean ($\mu$) and Std. Dev. ($\sigma$) values for the 20 runs. We also provide the $TIRR_{ft}$ metric and the number of passengers in the test input required to reproduce the failure}
\label{table:TimeResults}
\resizebox{\textwidth}{!}{
\begin{tabular}{llrrrrr|rrrrr|rrrrr} 
\cline{3-17}
                                       &                                     & \multicolumn{5}{c}{\textbf{Test input 1}}                                                                                              & \multicolumn{5}{c}{\textbf{Test input 2}}                                                                                                        & \multicolumn{5}{c}{\textbf{Test input 3}}                                                                                                         \\ 
\hline
\multirow{2}{*}{\textbf{Threshold \%}} & \multirow{2}{*}{\textbf{Algorithm}} & \multicolumn{3}{c}{\textbf{Execution Time}}       & \multirow{2}{*}{\textbf{$TIRR_{ft}$}} & \multicolumn{1}{c|}{\textbf{\# of~}} & \multicolumn{3}{c}{\textbf{Execution Time}}       & \multirow{2}{*}{\textbf{$TIRR_{ft}$}} & \multicolumn{1}{c|}{\textbf{\textbf{\# of~}}}  & \multicolumn{3}{c}{\textbf{Execution Time}}       & \multirow{2}{*}{\textbf{$TIRR_{ft}$}} & \multicolumn{1}{c}{\textbf{\textbf{\# of~}}}    \\ 
\cline{3-5}\cline{8-10}\cline{13-15}
                                       &                                     & \textbf{$m$} & \textbf{$\mu$} & \textbf{$\sigma$} &                                       & \textbf{\textbf{Passengers}}         & \textbf{$m$} & \textbf{$\mu$} & \textbf{$\sigma$} &                                       & \textbf{\textbf{\textbf{\textbf{Passengers}}}} & \textbf{$m$} & \textbf{$\mu$} & \textbf{$\sigma$} &                                       & \textbf{\textbf{\textbf{\textbf{Passengers}}}}  \\ 
\hline
                                       & $DD_{EB}$                             & 664.19       & 566.50         & 206.23            & 0.55                                  & 1068                                 & 423.61       & 394.93         & 82.86             & 0.93                                  & 111                                            & 472.08       & 432.81         & 103.09            & 0.96                                  & 88                                              \\
                                       & $DD_{TB}$                             & 572.92       & 574.25         & 14.98             & 0.69                                  & 742                                  & 418.67       & 417.90         & 11.26             & 0.95                                  & 82                                             & 390.57       & 393.32         & 11.19             & 0.96                                  & 91                                              \\
5\%                                    & $EWDD_{EB}$                           & 375.27       & 367.65         & 29.15             & 0.86                                  & 340                                  & 290.37       & 290.28         & 11.02             & 0.95                                  & 69                                             & 280.77       & 281.09         & 11.71             & 0.98                                  & 54                                              \\
                                       & $EWDD_{TB}$                           & 314.80       & 316.25         & 7.88              & 0.86                                  & 340                                  & 286.86       & 282.88         & 9.80              & 0.95                                  & 72                                             & 255.39       & 256.25         & 10.00             & 0.98                                  & 54                                              \\
                                       & $Backward$                          & 5464.74      & 5462.44        & 27.02             & 0.93                                  & 164                                  & 2793.22      & 2794.93        & 5.67              & 0.95                                  & 82                                             & 2992.67      & 2989.61        & 12.24             & 0.96                                  & 88                                              \\ 
\hline
                                       & $DD_{EB}$                             & 632.53       & 565.25         & 138.34            & 0.55                                  & 1068                                 & 415.18       & 394.86         & 57.35             & 0.93                                  & 111                                            & 451.63       & 432.39         & 44.70             & 0.97                                  & 72                                              \\
                                       & $DD_{TB}$                             & 572.23       & 574.32         & 14.30             & 0.69                                  & 742                                  & 442.02       & 440.94         & 11.09             & 0.95                                  & 71                                             & 390.76       & 392.76         & 12.28             & 0.97                                  & 76                                              \\
7\%                                    & $EWDD_{EB}$                           & 364.16       & 362.01         & 17.57             & 0.86                                  & 340                                  & 290.30       & 290.03         & 12.82             & 0.95                                  & 69                                             & 281.15       & 279.06         & 13.37             & 0.98                                  & 54                                              \\
                                       & $EWDD_{TB}$                           & 314.80       & 316.86         & 10.39             & 0.86                                  & 340                                  & 285.46       & 281.63         & 12.13             & 0.95                                  & 72                                             & 253.98       & 255.77         & 8.34              & 0.98                                  & 54                                              \\
                                       & $Backward$                          & 5464.01      & 5465.59        & 15.03             & 0.93                                  & 164                                  & 2438.48      & 2440           & 7.39              & 0.95                                  & 71                                             & 2443.23      & 2441.55        & 8.66              & 0.97                                  & 71                                              \\ 
\hline
                                       & $DD_{EB}$                             & 614.33       & 560.50         & 91.56             & 0.55                                  & 1068                                 & 403.28       & 395.00         & 35.18             & 0.93                                  & 111                                            & 450.04       & 432.56         & 47.92             & 0.97                                  & 72                                              \\
                                       & $DD_{TB}$                             & 571.11       & 565.06         & 14.33             & 0.69                                  & 742                                  & 443.44       & 444.34         & 13.16             & 0.95                                  & 71                                             & 392.03       & 393.38         & 12.31             & 0.97                                  & 76                                              \\
9\%                                    & $EWDD_{EB}$                           & 371.18       & 363.41         & 25.78             & 0.86                                  & 340                                  & 291.18       & 289.36         & 15.34             & 0.95                                  & 69                                             & 284.28       & 282.22         & 12.50             & 0.98                                  & 54                                              \\
                                       & $EWDD_{TB}$                           & 315.17       & 315.67         & 7.40              & 0.86                                  & 340                                  & 263.18       & 259.05         & 7.61              & 0.98                                  & 36                                             & 254.39       & 254.89         & 9.64              & 0.98                                  & 54                                              \\
                                       & $Backward$                          & 5131.84      & 5130.11        & 8.65              & 0.94                                  & 154                                  & 2437.04      & 2438.54        & 4.62              & 0.95                                  & 71                                             & 2439.21      & 2440.49        & 4.31              & 0.97                                  & 71                                              \\ 
\hline
                                       & $DD_{EB}$                             & 612.44       & 565.10         & 90.26             & 0.55                                  & 1068                                 & 402.17       & 394.99         & 32.80             & 0.93                                  & 111                                            & 442.60       & 432.68         & 34.96             & 0.97                                  & 72                                              \\
                                       & $DD_{TB}$                             & 571.59       & 574.28         & 12.86             & 0.69                                  & 742                                  & 444.91       & 444.28         & 12.69             & 0.95                                  & 71                                             & 393.87       & 393.59         & 12.61             & 0.97                                  & 76                                              \\
10\%                                   & $EWDD_{EB}$                           & 370.89       & 359.21         & 24.40             & 0.86                                  & 340                                  & 288.91       & 289.46         & 12.17             & 0.95                                  & 69                                             & 279.59       & 278.79         & 11.37             & 0.98                                  & 54                                              \\
                                       & $EWDD_{TB}$                           & 311.69       & 314.48         & 8.42              & 0.86                                  & 340                                  & 265.10       & 266.99         & 11.47             & 0.98                                  & 36                                             & 256.45       & 256.62         & 11.47             & 0.98                                  & 54                                              \\
                                       & $Backward$                          & 5130.20      & 5129.17        & 7.86              & 0.94                                  & 154                                  & 2437.11      & 2438.56        & 5.99              & 0.95                                  & 71                                             & 2022.89      & 2021.30        & 8.51              & 0.97                                  & 58                                              \\ 
\hline
                                       & $DD_{EB}$                             & 604.57       & 565.30         & 82.33             & 0.55                                  & 1068                                 & 407.15       & 394.92         & 43.50             & 0.93                                  & 111                                            & 453.38       & 432.55         & 74.57             & 0.97                                  & 72                                              \\
                                       & $DD_{TB}$                             & 570.74       & 564.33         & 12.79             & 0.69                                  & 742                                  & 442.35       & 441.73         & 11.95             & 0.95                                  & 71                                             & 396.11       & 402.87         & 12.27             & 0.97                                  & 76                                              \\
11\%                                   & $EWDD_{EB}$                           & 364.95       & 358.30         & 19.63             & 0.86                                  & 340                                  & 291.46       & 292.12         & 13.70             & 0.95                                  & 69                                             & 284.13       & 284.41         & 14.33             & 0.98                                  & 54                                              \\
                                       & $EWDD_{TB}$                           & 315.79       & 316.78         & 10.19             & 0.86                                  & 340                                  & 262.50       & 258.94         & 9.24              & 0.98                                  & 36                                             & 252.50       & 253.22         & 8.93              & 0.98                                  & 54                                              \\
                                       & $Backward$                          & 4397.45      & 4394.10        & 21.76             & 0.94                                  & 131                                  & 2315.64      & 2312.03        & 8.30              & 0.96                                  & 67                                             & 2023.89      & 2021.79        & 5.93              & 0.97                                  & 58                                              \\ 
\hline
                                       & $DD_{EB}$                             & 508.76       & 450.63         & 122.58            & 0.83                                  & 397                                  & 427.12       & 394.87         & 95.02             & 0.93                                  & 111                                            & 469.55       & 432.74         & 93.03             & 0.97                                  & 72                                              \\
                                       & $DD_{TB}$                             & 496.27       & 493.48         & 11.75             & 0.76                                  & 560                                  & 444.02       & 446.78         & 11.80             & 0.95                                  & 71                                             & 394.51       & 394.20         & 12.88             & 0.97                                  & 76                                              \\
13\%                                   & $EWDD_{EB}$                           & 371.60       & 376.10         & 23.53             & 0.86                                  & 340                                  & 289.46       & 289.60         & 12.01             & 0.95                                  & 69                                             & 282.50       & 278.91         & 13.58             & 0.98                                  & 54                                              \\
                                       & $EWDD_{TB}$                           & 313.87       & 316.36         & 5.10              & 0.86                                  & 340                                  & 263.43       & 262.68         & 8.08              & 0.98                                  & 36                                             & 254.38       & 254.90         & 8.48              & 0.98                                  & 54                                              \\
                                       & $Backward$                          & 4393.13      & 4393.89        & 5.54              & 0.94                                  & 131                                  & 2248.95      & 2246.15        & 6.95              & 0.96                                  & 65                                             & 2025.15      & 2022.88        & 5.68              & 0.97                                  & 58                                              \\ 
\hline
                                       & $DD_{EB}$                             & 500.60       & 450.65         & 92.74             & 0.83                                  & 397                                  & 414.91       & 394.81         & 64.02             & 0.93                                  & 111                                            & 449.29       & 432.62         & 41.25             & 0.97                                  & 72                                              \\
                                       & $DD_{TB}$                             & 495.24       & 488.88         & 11.86             & 0.76                                  & 560                                  & 444.45       & 444.07         & 11.05             & 0.95                                  & 71                                             & 391.51       & 393.74         & 10.77             & 0.97                                  & 76                                              \\
15\%                                   & $EWDD_{EB}$                           & 367.84       & 361.76         & 22.85             & 0.86                                  & 340                                  & 290.52       & 289.52         & 12.88             & 0.95                                  & 69                                             & 281.13       & 279.44         & 13.65             & 0.98                                  & 54                                              \\
                                       & $EWDD_{TB}$                           & 312.72       & 312.00         & 10.75             & 0.86                                  & 340                                  & 260.24       & 256.59         & 11.89             & 0.98                                  & 36                                             & 254.27       & 255.79         & 9.67              & 0.98                                  & 54                                              \\
                                       & $Backward$                          & 4393.07      & 4393.20        & 6.79              & 0.94                                  & 131                                  & 2249.66      & 2247.10        & 9.50              & 0.96                                  & 65                                             & 2026.62      & 2022.51        & 10.18             & 0.97                                  & 58                                              \\ 
\hline
                                       & $DD_{EB}$                             & 495.28       & 455.62         & 73.91             & 0.83                                  & 397                                  & 403.81       & 394.89         & 40.62             & 0.93                                  & 111                                            & 443.47       & 432.44         & 33.98             & 0.97                                  & 72                                              \\
                                       & $DD_{TB}$                             & 496.06       & 498.89         & 10.65             & 0.76                                  & 560                                  & 446.12       & 448.28         & 11.30             & 0.95                                  & 71                                             & 394.41       & 393.61         & 13.09             & 0.97                                  & 76                                              \\
17\%                                   & $EWDD_{EB}$                           & 366.92       & 365.46         & 21.04             & 0.86                                  & 340                                  & 289.85       & 290.18         & 12.33             & 0.95                                  & 69                                             & 281.34       & 279.28         & 14.12             & 0.98                                  & 54                                              \\
                                       & $EWDD_{TB}$                           & 318.68       & 317.08         & 10.97             & 0.86                                  & 340                                  & 263.43       & 259.13         & 7.17              & 0.98                                  & 36                                             & 253.56       & 252.27         & 8.76              & 0.98                                  & 54                                              \\
                                       & $Backward$                          & 4400.05      & 4394.76        & 23.85             & 0.94                                  & 131                                  & 2250.86      & 2246.53        & 8.39              & 0.96                                  & 65                                             & 2024.91      & 2022.68        & 6.00              & 0.97                                  & 58                                              \\ 
\hline
\multirow{2}{*}{}                      & $DD_{EB}$                             & 485.30       & 450.58         & 63.15             & 0.83                                  & 397                                  & 397.64       & 394.17         & 29.24             & 0.96                                  & 65                                             & 443.11       & 432.84         & 38.19             & 0.97                                  & 72                                              \\
                                       & $DD_{TB}$                             & 497.03       & 499.43         & 11.47             & 0.76                                  & 560                                  & 442.69       & 442.12         & 9.93              & 0.95                                  & 71                                             & 392.58       & 393.49         & 10.91             & 0.97                                  & 76                                              \\
19\%                                   & $EWDD_{EB}$                           & 370.09       & 365.35         & 24.10             & 0.86                                  & 340                                  & 288.10       & 289.48         & 9.70              & 0.95                                  & 69                                             & 281.07       & 278.54         & 13.49             & 0.98                                  & 54                                              \\
\multirow{2}{*}{}                      & $EWDD_{TB}$                           & 313.03       & 314.89         & 12.07             & 0.86                                  & 340                                  & 264.77       & 265.02         & 11.72             & 0.98                                  & 36                                             & 254.19       & 255.27         & 9.21              & 0.98                                  & 54                                              \\
                                       & $Backward$                          & 4393.76      & 4394.42        & 8.30              & 0.94                                  & 131                                  & 2248.49      & 2247.06        & 4.28              & 0.96                                  & 65                                             & 2025.77      & 2022.13        & 7.80              & 0.97                                  & 58                                              \\ 
\hline
                                       & $DD_{EB}$                             & 485.88       & 460.42         & 62.11             & 0.83                                  & 397                                  & 412.19       & 394.10         & 65.67             & 0.96                                  & 65                                             & 468.87       & 432.62         & 90.92             & 0.97                                  & 72                                              \\
                                       & $DD_{TB}$                             & 496.87       & 494.18         & 12.79             & 0.76                                  & 560                                  & 442.50       & 443.74         & 10.51             & 0.95                                  & 69                                             & 390.82       & 388.77         & 12.47             & 0.97                                  & 76                                              \\
20\%                                   & $EWDD_{EB}$                           & 367.34       & 354.70         & 25.07             & 0.86                                  & 340                                  & 291.34       & 289.64         & 14.07             & 0.95                                  & 69                                             & 283.33       & 279.25         & 16.46             & 0.98                                  & 54                                              \\
                                       & $EWDD_{TB}$                           & 315.71       & 317.73         & 8.45              & 0.86                                  & 340                                  & 261.79       & 258.43         & 8.18              & 0.98                                  & 36                                             & 251.33       & 248.35         & 12.03             & 0.98                                  & 54                                              \\
                                       & $Backward$                          & 4391.59      & 4393.25        & 7.32              & 0.94                                  & 131                                  & 2249.84      & 2246.65        & 7.67              & 0.96                                  & 65                                             & 2027         & 2022.69        & 8.73              & 0.97                                  & 58                                              \\
\hline
\end{tabular}
}
\end{table*}

%% file: Tables/RQ3Table.tex
\begin{table}[ht]
\caption{RQ3 -- Summary of the statistical test results for the Execution Time metric}
\label{table:RQ3}
\centering

\resizebox{8cm}{!}{\begin{tabular}{lllllllll}
\cline{4-9}
 & \textbf{} & \textbf{} & \multicolumn{2}{c}{\textbf{Test input 1}} & \multicolumn{2}{c}{\textbf{Test input 2}} & \multicolumn{2}{c}{\textbf{Test input 3}} \\ \hline
\multicolumn{1}{c}{\textbf{Thres (\%)}} & \multicolumn{1}{c}{\textbf{Algorithm A}} & \multicolumn{1}{c}{\textbf{Algorithm B}} & \textbf{\^{A}$_{12}$} & \textbf{p-value} & \textbf{\^{A}$_{12}$} & \textbf{p-value} & \textbf{\^{A}$_{12}$} & \textbf{p-value} \\ \hline
\multirow{2}{*}{5\%} & DD\textsubscript{TB} & DD\textsubscript{EB} & 0.48 & 0.8392 & 0.75 & 0.0077 & 0.01 & \textless{}0.0001 \\
 & EWDD\textsubscript{TB} & EWDD\textsubscript{EB} & 0.00& \textless{}0.0001 & 0.45 & 0.5792 & 0.07 & \textless{}0.0001 \\ \hline
\multirow{2}{*}{7\%} & DD\textsubscript{TB} & DD\textsubscript{EB} & 0.53 & 0.7557 & 0.76 & 0.0055 & 0.01 & \textless{}0.0001 \\
 & EWDD\textsubscript{TB} & EWDD\textsubscript{EB} & 0.00& \textless{}0.0001 & 0.44 & 0.4248 & 0.02 & \textless{}0.0001 \\ \hline
\multirow{2}{*}{9\%} & DD\textsubscript{TB} & DD\textsubscript{EB} & 0.56 & 0.5075 & 0.87 & \textless{}0.0001 & 0.03 & \textless{}0.0001 \\
 & EWDD\textsubscript{TB} & EWDD\textsubscript{EB} & 0.00& \textless{}0.0001 & 0.02 & \textless{}0.0001 & 0.02 & \textless{}0.0001 \\ \hline
\multirow{2}{*}{10\%} & DD\textsubscript{TB} & DD\textsubscript{EB} & 0.58 & 0.4093 & 0.84 & 0.0002 & 0.05 & \textless{}0.0001 \\
 & EWDD\textsubscript{TB} & EWDD\textsubscript{EB} & 0.00& \textless{}0.0001 & 0.08 & \textless{}0.0001 & 0.07 & \textless{}0.0001 \\ \hline
\multirow{2}{*}{11\%} & DD\textsubscript{TB} & DD\textsubscript{EB} & 0.53 & 0.7557 & 0.83 & 0.0004 & 0.07 & \textless{}0.0001 \\
 & EWDD\textsubscript{TB} & EWDD\textsubscript{EB} & 0.00& \textless{}0.0001 & 0.05 & \textless{}0.0001 & 0.02 & \textless{}0.0001 \\ \hline
\multirow{2}{*}{13\%} & DD\textsubscript{TB} & DD\textsubscript{EB} & 0.68 & 0.0565 & 0.76 & 0.0060.00& 0.05 & \textless{}0.0001 \\
 & EWDD\textsubscript{TB} & EWDD\textsubscript{EB} & 0.00& \textless{}0.0001 & 0.03 & \textless{}0.0001 & 0.03 & \textless{}0.0001 \\ \hline
\multirow{2}{*}{15\%} & DD\textsubscript{TB} & DD\textsubscript{EB} & 0.68 & 0.0531 & 0.78 & 0.0023 & 0.03 & \textless{}0.0001 \\
 & EWDD\textsubscript{TB} & EWDD\textsubscript{EB} & 0.00& \textless{}0.0001 & 0.03 & \textless{}0.0001 & 0.04 & \textless{}0.0001 \\ \hline
\multirow{2}{*}{17\%} & DD\textsubscript{TB} & DD\textsubscript{EB} & 0.66 & 0.0962 & 0.80 & 0.0012 & 0.06 & \textless{}0.0001 \\
 & EWDD\textsubscript{TB} & EWDD\textsubscript{EB} & 0.00& \textless{}0.0001 & 0.03 & \textless{}0.0001 & 0.05 & \textless{}0.0001 \\ \hline
\multirow{2}{*}{19\%} & DD\textsubscript{TB} & DD\textsubscript{EB} & 0.73 & 0.0133 & 0.90 & \textless{}0.0001 & 0.06 & \textless{}0.0001 \\
 & EWDD\textsubscript{TB} & EWDD\textsubscript{EB} & 0.00& \textless{}0.0001 & 0.08 & \textless{}0.0001 & 0.04 & \textless{}0.0001 \\ \hline
\multirow{2}{*}{20\%} & DD\textsubscript{TB} & DD\textsubscript{EB} & 0.71 & 0.0256 & 0.82 & 0.0006 & 0.04 & \textless{}0.0001 \\
 & EWDD\textsubscript{TB} & EWDD\textsubscript{EB} & 0.00& \textless{}0.0001 & 0.02 & \textless{}0.0001 & 0.04 & \textless{}0.0001 \\ \cline{1-9} 
\end{tabular}}

\end{table}

%% file: lessons.tex
We now summarize the key lessons learned from the application and extension of the delta debugging algorithm in Orona for debugging their elevator dispatching algorithms:

\textbf{Lesson 1 -- Using information of the CPS environment is beneficial for isolating the failure-inducing test inputs:} One core contribution of this paper relates to the environment-wise delta debugging algorithm, which monitors the environment and the physical aspects of the CPS to identify static situations. This helps the delta debugging algorithm to be faster and to reduce the failure-inducing test inputs even more.

\textbf{Lesson 2 -- In some cases, the most experienced developers were able to identify the problem with only seeing the test input and output, without the need for executing any simulation:} When qualitatively analyzing the test inputs provided by the algorithms with practitioners, we saw that the two most experienced developers could identify the root cause of the failure without executing the test input in the simulator. Conversely, the less experienced ones still did not feel confident to guess a problem without executing a simulation. We believe this lessons learned could be eventually translated to other CPSs and simulators. For instance, test cases in MATLAB/Simulink, a well-known CPS simulator, is composed by signals over time. With our approach, minimized input and output signals could be recorded and provided to the engineer for inspection.

\textbf{Lesson 3 -- The application of Delta Debugging could be useful beyond debugging:} Two of the experts claimed that eventually, delta debugging could be used for other purposes apart from debugging. For instance, they claimed that critical regions could be identified and later used in \textit{Hardware-in-the-Loop (HiL)} test executions, where the test execution time is critical. This lesson can help other practitioners from the CPS domain where HiL test executions are critical (e.g., the satellite domain~\cite{shin2021uncertainty}).

\textbf{Lesson 4 -- Delta debugging was found useful for manually proposing a patch:} The domain experts claimed that delta debugging could help identify the root-cause of the failure, and this could easily help to propose a patch, which was, in these cases, a change in configuration parameters. Three of the four interviewed experts proposed the same patch, which we later confirmed whether this patch was correct. Misconfigurations in CPSs are common~\cite{valle2023automated,han2022control,wang2021exploratory,garcia2020comprehensive}, and therefore believe that this lesson can eventually be transferred to other case study systems.

\textbf{Lesson 5 -- Specifying static situations was easy even for an inexperienced user:} The specification of static situations was easy. In fact, this was proposed by an academic researcher without extensive domain knowledge. Domain experts were only necessary to confirm that the defined static situations made sense to them when explaining the environment-wise delta debugging algorithm. Static situations are common in other CPSs too. For instance, in the context of an autonomous drone, a static situation can be when this is landed and with the engines at 0 rpms. In a pick-and-place robot, a static stiuation could be when the robotic arm is stopped and the conveyor does not have any piece that needs to be picked and placed somewhere else.

%% file: relatedWork.tex

A system of elevators is one type of CPS. Verification of CPSs has gained important attention in the last few years from different perspectives, including test generation~\cite{menghi2020approximation,abdessalem2020automated,abdessalem2018testing,luo2021targeting,arcaini2021parameter,matinnejad2018test,matinnejad2016automated,humeniuk2022search}, regression test optimization~\cite{matinnejad2018test,shin2018test,arrieta2019search,arrieta2019pareto} and the test oracle problem~\cite{ayerdi2020qos,ayerdi2021generating,menghi2019generating,stocco2020misbehaviour}. Nevertheless, despite its importance, research in debugging CPSs is quite limited. Some works focused on localizing faults of CPSs modeled in MATLAB/Simulink~\cite{liu2017improving,liu2016localizing,Liu2016,Liu2018,bartocci2022search}. Deshmukh et al.~\cite{deshmukh2018parameter} focused on localizing faults caused by missconfigurations. Bartocci et al.~\cite{bartocci2021cpsdebug} proposed a method to automatically explain failures in CPSs modeled in Simulink. Unlike all these studies, our approach focuses on reducing failure-inducing test inputs. 

Since its proposal in the beginning of the 2000s, delta debugging~\cite{hildebrandt2000simplifying,zeller2002simplifying} has been applied into multiple application domains, including microservices~\cite{zhou2018delta}, compiler testing~\cite{chen2019history} and web applications~\cite{hammoudi2015use}. Multiple approaches have been aiming to improve the effectiveness and/or efficiency of delta debugging. For instance, Misherghi and Su~\cite{10.1145/1134285.1134307} proposed the hierarchical delta debugging (HDD) algorithm to improve the performance of the traditional Delta Debugging algorithm for structured inputs (e.g., XML files). Gopinath et al.~\cite{gopinath2020abstracting} aimed at abstracting the failure-inducing inputs in order to isolate the failure and characterize the circumstances in which the failure occurs. Unlike our approach, none of these approaches are explicitly designed to target CPSs.

Our recent study implemented and studied the delta debugging approach for CPSs modeled in Simulink~\cite{valletowards}. Specifically, we found that 
delta debugging can help reduce both the failure-inducing test inputs, as well as the test coverage. However, there are four key differences between both approaches~\cite{valletowards}: (1) we enhance the delta debugging algorithm by proposing the environment-wise delta debugging algorithm; (2) we apply our approach into an industrial case study; (3) we conduct an empirical evaluation with an industrial case study and measure both the efficiency and effectiveness of the approach; (4) lastly, we assess the opinion of practitioners about the provided outputs by the isolation algorithms.

%% file: conclusion.tex
We propose and extend the delta debugging algorithm to reduce failure-inducing test inputs of elevator dispatching algorithms. We apply these algorithms in the dispatching algorithms of Orona, one of the leading European elevators company. Our extended version of the delta debugging algorithm, named environment-wise delta debugging, takes the advantage of monitoring the environment of the system of elevators to search static situations and help the delta debugging. This permits the isolation process to be 1.3 to 1.8 times faster than the traditional delta debugging approach, while obtaining a larger reduction of the failure-inducing test inputs. A qualitative analysis with domain experts confirmed our approach helps them understand the failure and propose a patch. 


The current version of the algorithm has already been transferred to Orona. To this end, a friendly user-interface was developed. Orona's engineers make weekly use of the tool to help debug issues of their dispatching algorithms. One core limitation of our approach lies in the manual assessment of the test inputs to determine under which circumstances a test input fails. In the future, we would like to study learning techniques that would explain such circumstances. A potential approach could be to adapt \textit{ALHAZEN}~\cite{kampmann2020does} to our context. However, this encompasses several challenges. For instance, in our context, the system does not necessarily fail based solely on some inputs, but also based on the internal states of the system. Moreover, in our context, the notion of time is important. A hypothesis example for a failure in our context could lay as follows: ``\textit{When all elevators have more than 5 passengers and there are 3 subsequent calls within 15 seconds, the waiting time of passengers in floor 5 will exceed certain threshold}''. As a result, significant tailoring would be required to such a learning algorithm to consider both, internal states of the system and time constraints.


%% file: ISSTA_Main.bbl

\begin{thebibliography}{50}


\ifx \showCODEN    \undefined \def \showCODEN     #1{\unskip}     \fi
\ifx \showDOI      \undefined \def \showDOI       #1{#1}\fi
\ifx \showISBNx    \undefined \def \showISBNx     #1{\unskip}     \fi
\ifx \showISBNxiii \undefined \def \showISBNxiii  #1{\unskip}     \fi
\ifx \showISSN     \undefined \def \showISSN      #1{\unskip}     \fi
\ifx \showLCCN     \undefined \def \showLCCN      #1{\unskip}     \fi
\ifx \shownote     \undefined \def \shownote      #1{#1}          \fi
\ifx \showarticletitle \undefined \def \showarticletitle #1{#1}   \fi
\ifx \showURL      \undefined \def \showURL       {\relax}        \fi
\providecommand\bibfield[2]{#2}
\providecommand\bibinfo[2]{#2}
\providecommand\natexlab[1]{#1}
\providecommand\showeprint[2][]{arXiv:#2}

\bibitem[Abdessalem et~al\mbox{.}(2018)]%
        {abdessalem2018testing}
\bibfield{author}{\bibinfo{person}{Raja~Ben Abdessalem}, \bibinfo{person}{Shiva
  Nejati}, \bibinfo{person}{Lionel~C Briand}, {and} \bibinfo{person}{Thomas
  Stifter}.} \bibinfo{year}{2018}\natexlab{}.
\newblock \showarticletitle{Testing vision-based control systems using
  learnable evolutionary algorithms}. In \bibinfo{booktitle}{\emph{2018
  IEEE/ACM 40th International Conference on Software Engineering (ICSE)}}.
  IEEE, \bibinfo{pages}{1016--1026}.
\newblock


\bibitem[Abdessalem et~al\mbox{.}(2020)]%
        {abdessalem2020automated}
\bibfield{author}{\bibinfo{person}{Raja~Ben Abdessalem},
  \bibinfo{person}{Annibale Panichella}, \bibinfo{person}{Shiva Nejati},
  \bibinfo{person}{Lionel~C Briand}, {and} \bibinfo{person}{Thomas Stifter}.}
  \bibinfo{year}{2020}\natexlab{}.
\newblock \showarticletitle{Automated repair of feature interaction failures in
  automated driving systems}. In \bibinfo{booktitle}{\emph{Proceedings of the
  29th ACM SIGSOFT International Symposium on Software Testing and Analysis}}.
  \bibinfo{pages}{88--100}.
\newblock


\bibitem[Ali and Yue(2015)]%
        {ali2015u}
\bibfield{author}{\bibinfo{person}{Shaukat Ali} {and} \bibinfo{person}{Tao
  Yue}.} \bibinfo{year}{2015}\natexlab{}.
\newblock \showarticletitle{U-test: evolving, modelling and testing realistic
  uncertain behaviours of cyber-physical systems}. In
  \bibinfo{booktitle}{\emph{2015 IEEE 8th International Conference on Software
  Testing, Verification and Validation (ICST)}}. IEEE, \bibinfo{pages}{1--2}.
\newblock


\bibitem[Alur(2015)]%
        {alur2015principles}
\bibfield{author}{\bibinfo{person}{Rajeev Alur}.}
  \bibinfo{year}{2015}\natexlab{}.
\newblock \bibinfo{booktitle}{\emph{Principles of cyber-physical systems}}.
\newblock \bibinfo{publisher}{MIT press}.
\newblock


\bibitem[Arcaini et~al\mbox{.}(2021)]%
        {arcaini2021parameter}
\bibfield{author}{\bibinfo{person}{Paolo Arcaini}, \bibinfo{person}{Alessandro
  Cal{\`o}}, \bibinfo{person}{Fuyuki Ishikawa}, \bibinfo{person}{Thomas
  Laurent}, \bibinfo{person}{Xiao-Yi Zhang}, \bibinfo{person}{Shaukat Ali},
  \bibinfo{person}{Florian Hauer}, {and} \bibinfo{person}{Anthony Ventresque}.}
  \bibinfo{year}{2021}\natexlab{}.
\newblock \showarticletitle{Parameter-Based Testing and Debugging of Autonomous
  Driving Systems}. In \bibinfo{booktitle}{\emph{2021 IEEE Intelligent Vehicles
  Symposium Workshops (IV Workshops)}}. IEEE, \bibinfo{pages}{197--202}.
\newblock


\bibitem[Arrieta et~al\mbox{.}(2022)]%
        {arrieta2022automating}
\bibfield{author}{\bibinfo{person}{Aitor Arrieta}, \bibinfo{person}{Maialen
  Otaegi}, \bibinfo{person}{Liping Han}, \bibinfo{person}{Goiuria Sagardui},
  \bibinfo{person}{Shaukat Ali}, {and} \bibinfo{person}{Maite Arratibel}.}
  \bibinfo{year}{2022}\natexlab{}.
\newblock \showarticletitle{Automating Test Oracle Generation in DevOps for
  Industrial Elevators}. In \bibinfo{booktitle}{\emph{2022 IEEE International
  Conference on Software Analysis, Evolution and Reengineering (SANER)}}. IEEE,
  \bibinfo{pages}{284--288}.
\newblock


\bibitem[Arrieta et~al\mbox{.}(2023)]%
        {arrieta2023some}
\bibfield{author}{\bibinfo{person}{Aitor Arrieta}, \bibinfo{person}{Pablo
  Valle}, \bibinfo{person}{Joseba~A Agirre}, {and} \bibinfo{person}{Goiuria
  Sagardui}.} \bibinfo{year}{2023}\natexlab{}.
\newblock \showarticletitle{Some seeds are strong: Seeding strategies for
  search-based test case selection}.
\newblock \bibinfo{journal}{\emph{ACM Transactions on Software Engineering and
  Methodology}} \bibinfo{volume}{32}, \bibinfo{number}{1}
  (\bibinfo{year}{2023}), \bibinfo{pages}{1--47}.
\newblock


\bibitem[Arrieta et~al\mbox{.}(2019a)]%
        {arrieta2019pareto}
\bibfield{author}{\bibinfo{person}{Aitor Arrieta}, \bibinfo{person}{Shuai
  Wang}, \bibinfo{person}{Urtzi Markiegi}, \bibinfo{person}{Ainhoa
  Arruabarrena}, \bibinfo{person}{Leire Etxeberria}, {and}
  \bibinfo{person}{Goiuria Sagardui}.} \bibinfo{year}{2019}\natexlab{a}.
\newblock \showarticletitle{Pareto efficient multi-objective black-box test
  case selection for simulation-based testing}.
\newblock \bibinfo{journal}{\emph{Information and Software Technology}}
  \bibinfo{volume}{114} (\bibinfo{year}{2019}), \bibinfo{pages}{137--154}.
\newblock


\bibitem[Arrieta et~al\mbox{.}(2017)]%
        {arrieta2017employing}
\bibfield{author}{\bibinfo{person}{Aitor Arrieta}, \bibinfo{person}{Shuai
  Wang}, \bibinfo{person}{Urtzi Markiegi}, \bibinfo{person}{Goiuria Sagardui},
  {and} \bibinfo{person}{Leire Etxeberria}.} \bibinfo{year}{2017}\natexlab{}.
\newblock \showarticletitle{Employing multi-objective search to enhance
  reactive test case generation and prioritization for testing industrial
  cyber-physical systems}.
\newblock \bibinfo{journal}{\emph{IEEE Transactions on Industrial Informatics}}
  \bibinfo{volume}{14}, \bibinfo{number}{3} (\bibinfo{year}{2017}),
  \bibinfo{pages}{1055--1066}.
\newblock


\bibitem[Arrieta et~al\mbox{.}(2016)]%
        {arrieta2016test}
\bibfield{author}{\bibinfo{person}{Aitor Arrieta}, \bibinfo{person}{Shuai
  Wang}, \bibinfo{person}{Goiuria Sagardui}, {and} \bibinfo{person}{Leire
  Etxeberria}.} \bibinfo{year}{2016}\natexlab{}.
\newblock \showarticletitle{Test case prioritization of configurable
  cyber-physical systems with weight-based search algorithms}. In
  \bibinfo{booktitle}{\emph{Proceedings of the Genetic and Evolutionary
  Computation Conference 2016}}. \bibinfo{pages}{1053--1060}.
\newblock


\bibitem[Arrieta et~al\mbox{.}(2019b)]%
        {arrieta2019search}
\bibfield{author}{\bibinfo{person}{Aitor Arrieta}, \bibinfo{person}{Shuai
  Wang}, \bibinfo{person}{Goiuria Sagardui}, {and} \bibinfo{person}{Leire
  Etxeberria}.} \bibinfo{year}{2019}\natexlab{b}.
\newblock \showarticletitle{Search-based test case prioritization for
  simulation-based testing of cyber-physical system product lines}.
\newblock \bibinfo{journal}{\emph{Journal of Systems and Software}}
  \bibinfo{volume}{149} (\bibinfo{year}{2019}), \bibinfo{pages}{1--34}.
\newblock


\bibitem[Ayerdi et~al\mbox{.}(2020a)]%
        {ayerdi2020towards}
\bibfield{author}{\bibinfo{person}{Jon Ayerdi}, \bibinfo{person}{Aitor
  Garciandia}, \bibinfo{person}{Aitor Arrieta}, \bibinfo{person}{Wasif Afzal},
  \bibinfo{person}{Eduard Enoiu}, \bibinfo{person}{Aitor Agirre},
  \bibinfo{person}{Goiuria Sagardui}, \bibinfo{person}{Maite Arratibel}, {and}
  \bibinfo{person}{Ola Sellin}.} \bibinfo{year}{2020}\natexlab{a}.
\newblock \showarticletitle{Towards a taxonomy for eliciting design-operation
  continuum requirements of cyber-physical systems}. In
  \bibinfo{booktitle}{\emph{2020 IEEE 28th International Requirements
  Engineering Conference (RE)}}. IEEE, \bibinfo{pages}{280--290}.
\newblock


\bibitem[Ayerdi et~al\mbox{.}(2020b)]%
        {ayerdi2020qos}
\bibfield{author}{\bibinfo{person}{Jon Ayerdi}, \bibinfo{person}{Sergio
  Segura}, \bibinfo{person}{Aitor Arrieta}, \bibinfo{person}{Goiuria Sagardui},
  {and} \bibinfo{person}{Maite Arratibel}.} \bibinfo{year}{2020}\natexlab{b}.
\newblock \showarticletitle{Qos-aware metamorphic testing: An elevation case
  study}. In \bibinfo{booktitle}{\emph{2020 IEEE 31st International Symposium
  on Software Reliability Engineering (ISSRE)}}. IEEE,
  \bibinfo{pages}{104--114}.
\newblock


\bibitem[Ayerdi et~al\mbox{.}(2021)]%
        {ayerdi2021generating}
\bibfield{author}{\bibinfo{person}{Jon Ayerdi}, \bibinfo{person}{Valerio
  Terragni}, \bibinfo{person}{Aitor Arrieta}, \bibinfo{person}{Paolo Tonella},
  \bibinfo{person}{Goiuria Sagardui}, {and} \bibinfo{person}{Maite Arratibel}.}
  \bibinfo{year}{2021}\natexlab{}.
\newblock \showarticletitle{Generating metamorphic relations for cyber-physical
  systems with genetic programming: an industrial case study}. In
  \bibinfo{booktitle}{\emph{Proceedings of the 29th ACM Joint Meeting on
  European Software Engineering Conference and Symposium on the Foundations of
  Software Engineering}}. \bibinfo{pages}{1264--1274}.
\newblock


\bibitem[Baheti and Gill(2011)]%
        {baheti2011cyber}
\bibfield{author}{\bibinfo{person}{Radhakisan Baheti} {and}
  \bibinfo{person}{Helen Gill}.} \bibinfo{year}{2011}\natexlab{}.
\newblock \showarticletitle{Cyber-physical systems}.
\newblock \bibinfo{journal}{\emph{The impact of control technology}}
  \bibinfo{volume}{12}, \bibinfo{number}{1} (\bibinfo{year}{2011}),
  \bibinfo{pages}{161--166}.
\newblock


\bibitem[Barney and Al-Sharif(2015)]%
        {barney2015elevator}
\bibfield{author}{\bibinfo{person}{Gina Barney} {and} \bibinfo{person}{Lutfi
  Al-Sharif}.} \bibinfo{year}{2015}\natexlab{}.
\newblock \bibinfo{booktitle}{\emph{Elevator traffic handbook: theory and
  practice}}.
\newblock \bibinfo{publisher}{Routledge}.
\newblock


\bibitem[Bartocci et~al\mbox{.}(2021)]%
        {bartocci2021cpsdebug}
\bibfield{author}{\bibinfo{person}{Ezio Bartocci}, \bibinfo{person}{Niveditha
  Manjunath}, \bibinfo{person}{Leonardo Mariani}, \bibinfo{person}{Cristinel
  Mateis}, {and} \bibinfo{person}{Dejan Ni{\v{c}}kovi{\'c}}.}
  \bibinfo{year}{2021}\natexlab{}.
\newblock \showarticletitle{CPSDebug: Automatic failure explanation in CPS
  models}.
\newblock \bibinfo{journal}{\emph{International Journal on Software Tools for
  Technology Transfer}} (\bibinfo{year}{2021}), \bibinfo{pages}{1--14}.
\newblock


\bibitem[Bartocci et~al\mbox{.}(2022)]%
        {bartocci2022search}
\bibfield{author}{\bibinfo{person}{Ezio Bartocci}, \bibinfo{person}{Leonardo
  Mariani}, \bibinfo{person}{Dejan Ni{\v{c}}kovi{\'c}}, {and}
  \bibinfo{person}{Drishti Yadav}.} \bibinfo{year}{2022}\natexlab{}.
\newblock \showarticletitle{Search-based Testing for Accurate Fault
  Localization in CPS}. In \bibinfo{booktitle}{\emph{2022 IEEE 33rd
  International Symposium on Software Reliability Engineering (ISSRE)}}. IEEE,
  \bibinfo{pages}{145--156}.
\newblock


\bibitem[Chen et~al\mbox{.}(2019)]%
        {chen2019history}
\bibfield{author}{\bibinfo{person}{Junjie Chen}, \bibinfo{person}{Guancheng
  Wang}, \bibinfo{person}{Dan Hao}, \bibinfo{person}{Yingfei Xiong},
  \bibinfo{person}{Hongyu Zhang}, {and} \bibinfo{person}{Lu Zhang}.}
  \bibinfo{year}{2019}\natexlab{}.
\newblock \showarticletitle{History-guided configuration diversification for
  compiler test-program generation}. In \bibinfo{booktitle}{\emph{2019 34th
  IEEE/ACM International Conference on Automated Software Engineering (ASE)}}.
  IEEE, \bibinfo{pages}{305--316}.
\newblock


\bibitem[Derler et~al\mbox{.}(2011)]%
        {derler2011modeling}
\bibfield{author}{\bibinfo{person}{Patricia Derler}, \bibinfo{person}{Edward~A
  Lee}, {and} \bibinfo{person}{Alberto~Sangiovanni Vincentelli}.}
  \bibinfo{year}{2011}\natexlab{}.
\newblock \showarticletitle{Modeling cyber--physical systems}.
\newblock \bibinfo{journal}{\emph{Proc. IEEE}} \bibinfo{volume}{100},
  \bibinfo{number}{1} (\bibinfo{year}{2011}), \bibinfo{pages}{13--28}.
\newblock


\bibitem[Deshmukh et~al\mbox{.}(2018)]%
        {deshmukh2018parameter}
\bibfield{author}{\bibinfo{person}{Jyotirmoy Deshmukh},
  \bibinfo{person}{Xiaoqing Jin}, \bibinfo{person}{Rupak Majumdar}, {and}
  \bibinfo{person}{Vinayak Prabhu}.} \bibinfo{year}{2018}\natexlab{}.
\newblock \showarticletitle{Parameter optimization in control software using
  statistical fault localization techniques}. In \bibinfo{booktitle}{\emph{2018
  ACM/IEEE 9th International Conference on Cyber-Physical Systems (ICCPS)}}.
  IEEE, \bibinfo{pages}{220--231}.
\newblock


\bibitem[{Garcia, Joshua and Feng, Yang and Shen, Junjie and Almanee, Sumaya
  and Xia, Yuan and Chen, and Qi Alfred}(2020)]%
        {garcia2020comprehensive}
\bibfield{author}{\bibinfo{person}{{Garcia, Joshua and Feng, Yang and Shen,
  Junjie and Almanee, Sumaya and Xia, Yuan and Chen, and Qi Alfred}}.}
  \bibinfo{year}{2020}\natexlab{}.
\newblock \showarticletitle{A comprehensive study of autonomous vehicle bugs}.
  In \bibinfo{booktitle}{\emph{Proceedings of the ACM/IEEE 42nd international
  conference on software engineering}}. \bibinfo{pages}{385--396}.
\newblock


\bibitem[Gopinath et~al\mbox{.}(2020)]%
        {gopinath2020abstracting}
\bibfield{author}{\bibinfo{person}{Rahul Gopinath}, \bibinfo{person}{Alexander
  Kampmann}, \bibinfo{person}{Nikolas Havrikov}, \bibinfo{person}{Ezekiel
  Soremekun}, {and} \bibinfo{person}{Andreas Zeller}.}
  \bibinfo{year}{2020}\natexlab{}.
\newblock \showarticletitle{Abstracting Failure-Inducing Inputs}. In
  \bibinfo{booktitle}{\emph{Proceedings of the 2020 International Symposium on
  Software Testing and Analysis}}. ACM.
\newblock


\bibitem[Hammoudi et~al\mbox{.}(2015)]%
        {hammoudi2015use}
\bibfield{author}{\bibinfo{person}{Mouna Hammoudi}, \bibinfo{person}{Brian
  Burg}, \bibinfo{person}{Gigon Bae}, {and} \bibinfo{person}{Gregg Rothermel}.}
  \bibinfo{year}{2015}\natexlab{}.
\newblock \showarticletitle{On the use of delta debugging to reduce recordings
  and facilitate debugging of web applications}. In
  \bibinfo{booktitle}{\emph{Proceedings of the 2015 10th Joint Meeting on
  Foundations of Software Engineering}}. \bibinfo{pages}{333--344}.
\newblock


\bibitem[Han et~al\mbox{.}(2022a)]%
        {han2022uncertainty}
\bibfield{author}{\bibinfo{person}{Liping Han}, \bibinfo{person}{Shaukat Ali},
  \bibinfo{person}{Tao Yue}, \bibinfo{person}{Aitor Arrieta}, {and}
  \bibinfo{person}{Maite Arratibel}.} \bibinfo{year}{2022}\natexlab{a}.
\newblock \showarticletitle{Uncertainty-aware Robustness Assessment of
  Industrial Elevator Systems}.
\newblock \bibinfo{journal}{\emph{ACM Transactions on Software Engineering and
  Methodology}} (\bibinfo{year}{2022}).
\newblock


\bibitem[Han et~al\mbox{.}(2022c)]%
        {han2022elevator}
\bibfield{author}{\bibinfo{person}{Liping Han}, \bibinfo{person}{Tao Yue},
  \bibinfo{person}{Shaukat Ali}, \bibinfo{person}{Aitor Arrieta}, {and}
  \bibinfo{person}{Maite Arratibel}.} \bibinfo{year}{2022}\natexlab{c}.
\newblock \showarticletitle{Are elevator software robust against uncertainties?
  results and experiences from an industrial case study}. In
  \bibinfo{booktitle}{\emph{Proceedings of the 30th ACM Joint European Software
  Engineering Conference and Symposium on the Foundations of Software
  Engineering}}. \bibinfo{pages}{1331--1342}.
\newblock


\bibitem[Han et~al\mbox{.}(2022b)]%
        {han2022control}
\bibfield{author}{\bibinfo{person}{Ruidong Han}, \bibinfo{person}{Chao Yang},
  \bibinfo{person}{Siqi Ma}, \bibinfo{person}{JiangFeng Ma},
  \bibinfo{person}{Cong Sun}, \bibinfo{person}{Juanru Li}, {and}
  \bibinfo{person}{Elisa Bertino}.} \bibinfo{year}{2022}\natexlab{b}.
\newblock \showarticletitle{Control parameters considered harmful: Detecting
  range specification bugs in drone configuration modules via learning-guided
  search}. In \bibinfo{booktitle}{\emph{Proceedings of the 44th International
  Conference on Software Engineering}}. \bibinfo{pages}{462--473}.
\newblock


\bibitem[Hildebrandt and Zeller(2000)]%
        {hildebrandt2000simplifying}
\bibfield{author}{\bibinfo{person}{Ralf Hildebrandt} {and}
  \bibinfo{person}{Andreas Zeller}.} \bibinfo{year}{2000}\natexlab{}.
\newblock \showarticletitle{Simplifying failure-inducing input}. In
  \bibinfo{booktitle}{\emph{Proceedings of the 2000 ACM SIGSOFT international
  symposium on Software testing and analysis}}. \bibinfo{pages}{135--145}.
\newblock


\bibitem[Humeniuk et~al\mbox{.}(2022)]%
        {humeniuk2022search}
\bibfield{author}{\bibinfo{person}{Dmytro Humeniuk}, \bibinfo{person}{Foutse
  Khomh}, {and} \bibinfo{person}{Giuliano Antoniol}.}
  \bibinfo{year}{2022}\natexlab{}.
\newblock \showarticletitle{A search-based framework for automatic generation
  of testing environments for cyber-physical systems}.
\newblock \bibinfo{journal}{\emph{Information and Software Technology}}
  (\bibinfo{year}{2022}), \bibinfo{pages}{106936}.
\newblock


\bibitem[Kampmann et~al\mbox{.}(2020)]%
        {kampmann2020does}
\bibfield{author}{\bibinfo{person}{Alexander Kampmann},
  \bibinfo{person}{Nikolas Havrikov}, \bibinfo{person}{Ezekiel~O Soremekun},
  {and} \bibinfo{person}{Andreas Zeller}.} \bibinfo{year}{2020}\natexlab{}.
\newblock \showarticletitle{When does my program do this? learning
  circumstances of software behavior}. In \bibinfo{booktitle}{\emph{Proceedings
  of the 28th ACM joint meeting on european software engineering conference and
  symposium on the foundations of software engineering}}.
  \bibinfo{pages}{1228--1239}.
\newblock


\bibitem[Liu et~al\mbox{.}(2016a)]%
        {Liu2016}
\bibfield{author}{\bibinfo{person}{Bing Liu}, \bibinfo{person}{Lucia},
  \bibinfo{person}{Shiva Nejati}, \bibinfo{person}{Lionel~C Briand}, {and}
  \bibinfo{person}{Thomas Bruckmann}.} \bibinfo{year}{2016}\natexlab{a}.
\newblock \showarticletitle{Simulink fault localization: an iterative
  statistical debugging approach}.
\newblock \bibinfo{journal}{\emph{Software Testing, Verification and
  Reliability}} \bibinfo{volume}{26}, \bibinfo{number}{6}
  (\bibinfo{year}{2016}), \bibinfo{pages}{431--459}.
\newblock


\bibitem[Liu et~al\mbox{.}(2016b)]%
        {liu2016localizing}
\bibfield{author}{\bibinfo{person}{Bing Liu}, \bibinfo{person}{Shiva Nejati},
  \bibinfo{person}{Lionel Briand}, \bibinfo{person}{Thomas Bruckmann},
  {et~al\mbox{.}}} \bibinfo{year}{2016}\natexlab{b}.
\newblock \showarticletitle{Localizing multiple faults in simulink models}. In
  \bibinfo{booktitle}{\emph{2016 IEEE 23rd International Conference on Software
  Analysis, Evolution, and Reengineering (SANER)}}, Vol.~\bibinfo{volume}{1}.
  IEEE, \bibinfo{pages}{146--156}.
\newblock


\bibitem[Liu et~al\mbox{.}(2017)]%
        {liu2017improving}
\bibfield{author}{\bibinfo{person}{Bing Liu}, \bibinfo{person}{Shiva Nejati},
  \bibinfo{person}{Lionel~C Briand}, {et~al\mbox{.}}}
  \bibinfo{year}{2017}\natexlab{}.
\newblock \showarticletitle{Improving fault localization for Simulink models
  using search-based testing and prediction models}. In
  \bibinfo{booktitle}{\emph{2017 IEEE 24th International Conference on Software
  Analysis, Evolution and Reengineering (SANER)}}. IEEE,
  \bibinfo{pages}{359--370}.
\newblock


\bibitem[Liu et~al\mbox{.}(2018)]%
        {Liu2018}
\bibfield{author}{\bibinfo{person}{Bing Liu}, \bibinfo{person}{Shiva Nejati},
  \bibinfo{person}{Lionel~C Briand}, {et~al\mbox{.}}}
  \bibinfo{year}{2018}\natexlab{}.
\newblock \showarticletitle{Effective fault localization of automotive Simulink
  models: achieving the trade-off between test oracle effort and fault
  localization accuracy}.
\newblock \bibinfo{journal}{\emph{Empirical Software Engineering}}
  (\bibinfo{year}{2018}), \bibinfo{pages}{1--47}.
\newblock


\bibitem[Luo et~al\mbox{.}(2021)]%
        {luo2021targeting}
\bibfield{author}{\bibinfo{person}{Yixing Luo}, \bibinfo{person}{Xiao-Yi
  Zhang}, \bibinfo{person}{Paolo Arcaini}, \bibinfo{person}{Zhi Jin},
  \bibinfo{person}{Haiyan Zhao}, \bibinfo{person}{Fuyuki Ishikawa},
  \bibinfo{person}{Rongxin Wu}, {and} \bibinfo{person}{Tao Xie}.}
  \bibinfo{year}{2021}\natexlab{}.
\newblock \showarticletitle{Targeting Requirements Violations of Autonomous
  Driving Systems by Dynamic Evolutionary Search}. In
  \bibinfo{booktitle}{\emph{2021 36th IEEE/ACM International Conference on
  Automated Software Engineering (ASE)}}. IEEE, \bibinfo{pages}{279--291}.
\newblock


\bibitem[Matinnejad et~al\mbox{.}(2016)]%
        {matinnejad2016automated}
\bibfield{author}{\bibinfo{person}{Reza Matinnejad}, \bibinfo{person}{Shiva
  Nejati}, \bibinfo{person}{Lionel~C Briand}, {and} \bibinfo{person}{Thomas
  Bruckmann}.} \bibinfo{year}{2016}\natexlab{}.
\newblock \showarticletitle{Automated test suite generation for time-continuous
  simulink models}. In \bibinfo{booktitle}{\emph{proceedings of the 38th
  International Conference on Software Engineering}}.
  \bibinfo{pages}{595--606}.
\newblock


\bibitem[Matinnejad et~al\mbox{.}(2018)]%
        {matinnejad2018test}
\bibfield{author}{\bibinfo{person}{Reza Matinnejad}, \bibinfo{person}{Shiva
  Nejati}, \bibinfo{person}{Lionel~C Briand}, {and} \bibinfo{person}{Thomas
  Bruckmann}.} \bibinfo{year}{2018}\natexlab{}.
\newblock \showarticletitle{Test generation and test prioritization for
  simulink models with dynamic behavior}.
\newblock \bibinfo{journal}{\emph{IEEE Transactions on Software Engineering}}
  \bibinfo{volume}{45}, \bibinfo{number}{9} (\bibinfo{year}{2018}),
  \bibinfo{pages}{919--944}.
\newblock


\bibitem[Menghi et~al\mbox{.}(2020)]%
        {menghi2020approximation}
\bibfield{author}{\bibinfo{person}{Claudio Menghi}, \bibinfo{person}{Shiva
  Nejati}, \bibinfo{person}{Lionel Briand}, {and} \bibinfo{person}{Yago~Isasi
  Parache}.} \bibinfo{year}{2020}\natexlab{}.
\newblock \showarticletitle{Approximation-refinement testing of
  compute-intensive cyber-physical models: An approach based on system
  identification}. In \bibinfo{booktitle}{\emph{2020 IEEE/ACM 42nd
  International Conference on Software Engineering (ICSE)}}. IEEE,
  \bibinfo{pages}{372--384}.
\newblock


\bibitem[Menghi et~al\mbox{.}(2019)]%
        {menghi2019generating}
\bibfield{author}{\bibinfo{person}{Claudio Menghi}, \bibinfo{person}{Shiva
  Nejati}, \bibinfo{person}{Khouloud Gaaloul}, {and} \bibinfo{person}{Lionel~C
  Briand}.} \bibinfo{year}{2019}\natexlab{}.
\newblock \showarticletitle{Generating automated and online test oracles for
  simulink models with continuous and uncertain behaviors}. In
  \bibinfo{booktitle}{\emph{Proceedings of the 2019 27th acm joint meeting on
  european software engineering conference and symposium on the foundations of
  software engineering}}. \bibinfo{pages}{27--38}.
\newblock


\bibitem[Misherghi and Su(2006)]%
        {10.1145/1134285.1134307}
\bibfield{author}{\bibinfo{person}{Ghassan Misherghi} {and}
  \bibinfo{person}{Zhendong Su}.} \bibinfo{year}{2006}\natexlab{}.
\newblock \showarticletitle{HDD: Hierarchical Delta Debugging}. In
  \bibinfo{booktitle}{\emph{Proceedings of the 28th International Conference on
  Software Engineering}} (Shanghai, China) \emph{(\bibinfo{series}{ICSE '06})}.
  \bibinfo{publisher}{Association for Computing Machinery},
  \bibinfo{address}{New York, NY, USA}, \bibinfo{pages}{142–151}.
\newblock
\showISBNx{1595933751}
\urldef\tempurl%
\url{https://doi.org/10.1145/1134285.1134307}
\showDOI{\tempurl}


\bibitem[Shin et~al\mbox{.}(2021)]%
        {shin2021uncertainty}
\bibfield{author}{\bibinfo{person}{Seung~Yeob Shin}, \bibinfo{person}{Karim
  Chaouch}, \bibinfo{person}{Shiva Nejati}, \bibinfo{person}{Mehrdad
  Sabetzadeh}, \bibinfo{person}{Lionel~C Briand}, {and} \bibinfo{person}{Frank
  Zimmer}.} \bibinfo{year}{2021}\natexlab{}.
\newblock \showarticletitle{Uncertainty-aware specification and analysis for
  hardware-in-the-loop testing of cyber-physical systems}.
\newblock \bibinfo{journal}{\emph{Journal of Systems and Software}}
  \bibinfo{volume}{171} (\bibinfo{year}{2021}), \bibinfo{pages}{110813}.
\newblock


\bibitem[Shin et~al\mbox{.}(2018)]%
        {shin2018test}
\bibfield{author}{\bibinfo{person}{Seung~Yeob Shin}, \bibinfo{person}{Shiva
  Nejati}, \bibinfo{person}{Mehrdad Sabetzadeh}, \bibinfo{person}{Lionel~C
  Briand}, {and} \bibinfo{person}{Frank Zimmer}.}
  \bibinfo{year}{2018}\natexlab{}.
\newblock \showarticletitle{Test case prioritization for acceptance testing of
  cyber physical systems: a multi-objective search-based approach}. In
  \bibinfo{booktitle}{\emph{Proceedings of the 27th acm sigsoft international
  symposium on software testing and analysis}}. \bibinfo{pages}{49--60}.
\newblock


\bibitem[Stocco et~al\mbox{.}(2020)]%
        {stocco2020misbehaviour}
\bibfield{author}{\bibinfo{person}{Andrea Stocco}, \bibinfo{person}{Michael
  Weiss}, \bibinfo{person}{Marco Calzana}, {and} \bibinfo{person}{Paolo
  Tonella}.} \bibinfo{year}{2020}\natexlab{}.
\newblock \showarticletitle{Misbehaviour prediction for autonomous driving
  systems}. In \bibinfo{booktitle}{\emph{Proceedings of the ACM/IEEE 42nd
  international conference on software engineering}}.
  \bibinfo{pages}{359--371}.
\newblock


\bibitem[Valle and Arrieta(2022)]%
        {valletowards}
\bibfield{author}{\bibinfo{person}{Pablo Valle} {and} \bibinfo{person}{Aitor
  Arrieta}.} \bibinfo{year}{2022}\natexlab{}.
\newblock \showarticletitle{Towards the Isolation of Failure-Inducing Inputs in
  Cyber-Physical Systems: is Delta Debugging Enough?}. In
  \bibinfo{booktitle}{\emph{2022 IEEE 29th International Conference on Software
  Analysis, Evolution and Reengineering (SANER)}}. IEEE,
  \bibinfo{pages}{549--553}.
\newblock


\bibitem[Valle et~al\mbox{.}(2023)]%
        {valle2023automated}
\bibfield{author}{\bibinfo{person}{Pablo Valle}, \bibinfo{person}{Aitor
  Arrieta}, {and} \bibinfo{person}{Maite Arratibel}.}
  \bibinfo{year}{2023}\natexlab{}.
\newblock \showarticletitle{Automated Misconfiguration Repair of Configurable
  Cyber-Physical Systems with Search: an Industrial Case Study on Elevator
  Dispatching Algorithms}. In \bibinfo{booktitle}{\emph{2023 IEEE/ACM 45th
  International Conference on Software Engineering (ICSE)}}.
  \bibinfo{pages}{396--408}.
\newblock


\bibitem[Wang et~al\mbox{.}(2021)]%
        {wang2021exploratory}
\bibfield{author}{\bibinfo{person}{Dinghua Wang}, \bibinfo{person}{Shuqing Li},
  \bibinfo{person}{Guanping Xiao}, \bibinfo{person}{Yepang Liu}, {and}
  \bibinfo{person}{Yulei Sui}.} \bibinfo{year}{2021}\natexlab{}.
\newblock \showarticletitle{An exploratory study of autopilot software bugs in
  unmanned aerial vehicles}. In \bibinfo{booktitle}{\emph{Proceedings of the
  29th ACM Joint Meeting on European Software Engineering Conference and
  Symposium on the Foundations of Software Engineering}}.
  \bibinfo{pages}{20--31}.
\newblock


\bibitem[Zeller and Hildebrandt(2002)]%
        {zeller2002simplifying}
\bibfield{author}{\bibinfo{person}{Andreas Zeller} {and} \bibinfo{person}{Ralf
  Hildebrandt}.} \bibinfo{year}{2002}\natexlab{}.
\newblock \showarticletitle{Simplifying and isolating failure-inducing input}.
\newblock \bibinfo{journal}{\emph{IEEE Transactions on Software Engineering}}
  \bibinfo{volume}{28}, \bibinfo{number}{2} (\bibinfo{year}{2002}),
  \bibinfo{pages}{183--200}.
\newblock


\bibitem[Zhang et~al\mbox{.}(2019)]%
        {zhang2019uncertainty}
\bibfield{author}{\bibinfo{person}{Man Zhang}, \bibinfo{person}{Shaukat Ali},
  \bibinfo{person}{Tao Yue}, \bibinfo{person}{Roland Norgren}, {and}
  \bibinfo{person}{Oscar Okariz}.} \bibinfo{year}{2019}\natexlab{}.
\newblock \showarticletitle{Uncertainty-wise cyber-physical system test
  modeling}.
\newblock \bibinfo{journal}{\emph{Software \& Systems Modeling}}
  \bibinfo{volume}{18}, \bibinfo{number}{2} (\bibinfo{year}{2019}),
  \bibinfo{pages}{1379--1418}.
\newblock


\bibitem[Zhang et~al\mbox{.}(2016)]%
        {zhang2016understanding}
\bibfield{author}{\bibinfo{person}{Man Zhang}, \bibinfo{person}{Bran Selic},
  \bibinfo{person}{Shaukat Ali}, \bibinfo{person}{Tao Yue},
  \bibinfo{person}{Oscar Okariz}, {and} \bibinfo{person}{Roland Norgren}.}
  \bibinfo{year}{2016}\natexlab{}.
\newblock \showarticletitle{Understanding uncertainty in cyber-physical
  systems: a conceptual model}. In \bibinfo{booktitle}{\emph{European
  conference on modelling foundations and applications}}. Springer,
  \bibinfo{pages}{247--264}.
\newblock


\bibitem[Zhou et~al\mbox{.}(2018)]%
        {zhou2018delta}
\bibfield{author}{\bibinfo{person}{Xiang Zhou}, \bibinfo{person}{Xin Peng},
  \bibinfo{person}{Tao Xie}, \bibinfo{person}{Jun Sun}, \bibinfo{person}{Wenhai
  Li}, \bibinfo{person}{Chao Ji}, {and} \bibinfo{person}{Dan Ding}.}
  \bibinfo{year}{2018}\natexlab{}.
\newblock \showarticletitle{Delta debugging microservice systems}. In
  \bibinfo{booktitle}{\emph{2018 33rd IEEE/ACM International Conference on
  Automated Software Engineering (ASE)}}. IEEE, \bibinfo{pages}{802--807}.
\newblock


\end{thebibliography}
